\documentclass[final,5p,times,twocolumn,authoryear,linenumbers]{elsarticle}
\usepackage{graphicx}
\usepackage{multirow}
\usepackage{txfonts}
\usepackage{chngcntr}
\usepackage{natbib}
\usepackage{url}
\usepackage{hyperref}
\usepackage{longtable}
\usepackage{tabularx}
\usepackage{amssymb}
\usepackage{lipsum}
\usepackage{array}
\usepackage{rotating}
\usepackage{changepage}
\usepackage{lineno}

\journal{Journal of High Energy Astrophysics}

\begin{document}

\begin{frontmatter}

\title{Revisiting the 2021 Outburst of the BHC MAXI J1803-298 Using NICER, NuSTAR, and Insight-HXMT Data}

\author[first,second]{Kaushik Chatterjee}
\affiliation[first]{organization={South-Western Institute for Astronomy Research, Yunnan University},
            addressline={University Town, Chenggong},
            city={Kunming},
            postcode={650500},
            state={Yunnan},
            country={People's Republic of China;
mails.kc.physics@gmail.com}}	
\affiliation[second]{Organization={Key Laboratory of Survey Science of Yunnan Province, Yunnan University}, 
	    city={Kunming}, 
	    state={Yunnan}, 
	    postcode={650500}, 
	    country={People's Republic of China;
kaushik@ynu.edu.cn}}

\author[third,fourth]{Sujoy K. Nath\fnref{equal}}
\affiliation[third]{organization={Indian Center for Space Physics},
            addressline={466 Barakhola, Netai Nagar},
            city={Kolkata},
            postcode={700099},
            state={West Bengal},
            country={India;
sujoynath0007@gmail.com}}
\affiliation[fourth]{organization={Institute of Astronomy, National Tsing Hua University},
	    city={Hsinchu},
	    postcode={300044},
	    country={Taiwan, ROC}}

\fntext[equal]{equal first author}

\begin{abstract}

We present a broadband spectral and timing study of the black hole candidate MAXI~J1803--298 during its 2021 outburst using simultaneous observations from NICER, \textit{NuSTAR}, 
and \textit{Insight-HXMT}.
The combined multi-instrument coverage allows us to investigate the evolution of low-frequency quasi-periodic oscillations (LFQPOs) together with the 
spectral properties of the source over a wide energy range.
During the early observation epoch, the source exhibits a hard or hard-intermediate spectral state dominated by 
Comptonized emission with reflection features.
Spectral modeling within the framework of the two-component advective flow (TCAF) model indicates the presence of a sub-Keplerian 
halo and a Keplerian disk with a shock located at $\sim$130 Schwarzschild radii, and provides an independent estimate of the black hole mass.
A prominent LFQPO is detected during 
this epoch with a centroid frequency evolving from $\sim$0.35~Hz to $\sim$0.5~Hz and extending up to $\sim$100~keV.
The energy-dependent fractional rms variability suggests that 
the modulation originates primarily from the Comptonizing inner accretion flow.
In contrast, a later observation epoch shows a softer spectral state characterized by stronger 
disk emission and a steeper photon index, during which no LFQPO is detected.
We also demonstrate that cospectral analysis effectively mitigates dead-time-induced distortions in 
\textit{NuSTAR} timing studies, confirming the intrinsic nature of the detected variability.
The combined spectral and timing results support a scenario in which LFQPOs in MAXI
J1803--298 arise from the dynamically evolving inner accretion flow.
\end{abstract}


\end{frontmatter}

\section{Introduction}

Black holes are among the simplest astronomical objects, yet still one of the most puzzling bodies out there in the universe.
These are the end products of massive stellar
bodies that ran out of their fuel.
Although their information is contained in three parameters: mass, spin, and charge, studying the radiation properties around the surroundings 
of these black holes reveals a great deal of insight into these systems.
Their radiation is mainly of two types: thermal and non-thermal radiation.
Matter from the companion can 
accrete onto these systems when it achieves critical viscosity and forms an accretion disk.
This has a Keplerian distribution of angular momentum. This is an optically thick and 
geometrically thin standard disk (Shakura \& Sunyaev 1973) that is responsible for the emission of thermal blackbody radiation by the conversion of gravitational potential energy 
into thermal energy as matter accretes inward (Frank, King \& Raine 1982).
Along with this, there is a hot Corona, which is the repository of hot electrons that intercept soft photons 
from the accretion disk and up-scatter them into higher energies in the process of inverse Comptonization (Sunyaev \& Titarchuk 1980; Chakrabarti \& Titarchuk 1995).
This producess 
the non-thermal part of the radiation. There could also be the production of jets/outflows, which are often thought to be launched from the hot Corona due to the thermal pressure 
(Chakrabarti 1998; Chakrabarti 1999).
This whole configuration controls the various kinds of observational features from black holes and their evolution with time.
A change in the physical configuration of the system 
changes the spectral nature as well as the outflowing jet properties.
Quasi-periodic oscillation (QPO) is an important property of black holes.
QPOs are associated with flux 
variability that is a direct consequence of the region from which they originate.
Generally, low-frequency QPOs (LFQPOs) are observed in stellar-mass black holes with the occasional 
occurrence of high-frequency QPOs (HFQPOs).
These could be associated with instabilities that produce the variability.
To explain the physical origin of LFQPOs in stellar-mass 
black holes, several theoretical frameworks have been put forward.
The relativistic precession model (Stella \& Vietri 1998) attributes QPO frequencies to the fundamental frequencies 
of particle orbits in the inner accretion disk, including nodal Lense-Thirring precession.
In the context of a truncated disk, LFQPOs are widely explained by the geometric Lense-Thirring 
precession of the entire hot inner accretion flow, which is misaligned with the black hole spin axis (Ingram, Done \& Fragile 2009).
Another proposed mechanism is the accretion-ejection 
instability model (Tagger \& Pellat 1999), which links LFQPOs to global spiral instabilities in the inner disk that transfer energy and angular momentum to a corona or jet.
Furthermore, 
within the two-component advective flow (TCAF) framework, oscillations of the post-shock Comptonizing region due to a resonance between the infall timescale of the post-shock matter 
and the cooling timescale are thought to produce the LFQPOs (Molteni, Sponholz \& Chakrabarti 1996; Chakrabarti et al. 2015).

The BHC MAXI J1803-298 was discovered on May 1, 2021. It was detected by the nova alert system using the gas slit camera (GSC) of the Monitor of All-sky X-ray Image (MAXI; 
Matsuoka et al. 2009) satellite.
A multi-wavelength follow-up of the source was presented by Sanchez (2022).
After its discovery, the outburst was monitored by various X-ray 
satellites, e.g., MAXI, NICER, AstroSat, NuSTAR, Insight-HXMT, etc. This system contains a rapidly spinning black hole with a spin of $a \sim 0.998$ (Coughenour et al. 2023).
The plane of the binary system is inclined at an angle of $i \sim (75 \pm 2)^\circ$ (Coughenour et al. 2023).
Follow-up optical (Buckley et al. 2021) and radio (Espinasse et 
al. 2021) monitoring of the source was reported.
Simultaneous spectroscopic analysis in both optical and X-ray (Bult et al. 2021; Homan et al. 2021; Xu \& Harrison 2021) suggested 
the source to be a BH.
The presence of disk wind signatures was also reported by these simultaneous spectroscopy studies (Miller \& Reynolds 2021; Sanchez et al. 2022).
When the 
outburst started, the source was in the HS during the rising phase.
Similar to a canonical outburst of a BH, this source has gone through every spectral state in the following 
sequence: HS (rising) $\rightarrow$ HIMS (rising) $\rightarrow$ SIMS $\rightarrow$ SS $\rightarrow$ SIMS $\rightarrow$ HIMS (declining) $\rightarrow$ HS (declining).
In the HS 
in the rising phase, before the source reached its peak, it showed periodic dips with a 7-hour period in data from observatories like NICER (Homan et al. 2021), NuSTAR (Xu \& 
Harrison 2021), and AstroSat (Jana et al. 2022).
QPOs were detected during the outburst in data from many satellites, such as NICER, NuSTAR, AstroSat, and Insight-HXMT.
The QPO 
frequency evolved from 0.13 Hz in the HS to 7.61 Hz in the intermediate state (Chand et al. 2021, 2022; Ubach et al. 2021).
Using AstroSat/LAXPC data, Jana et al. (2021) showed 
that the QPO feature stays prominent up to 30 keV.
From their spectral analysis, they reported a mass of the BH as $M_{BH} \sim$ 3.5-12.5 $M_\odot$.
Using AstroSat and NuSTAR 
light curve data in various energy bands, they estimated the period of the binary as $\sim$ 7 hour making the source one of the shortest-period binaries (Jana et al. 2022).
In this paper, we report on the timing and spectral analysis of the source using NICER, NuSTAR, and Insight-HXMT data.
Using AstroSat and NuSTAR light curve data in 
various energy bands, they estimated the period of the binary as $\sim$ 7 hour making the source one of the shortest-period binaries (Jana et al. 2022; Coughenour et al. 2023).
While previous studies have extensively documented the general outburst evolution, spectral state transitions, and orbital parameters of MAXI J1803--298 (e.g., Chand et al. 2022; 
Jana et al. 2022; Coughenour et al. 2023), the present study offers several novel advancements.
First, we perform broadband spectral modeling utilizing the physical two-component 
advective flow (TCAF) model for the first time on this source, yielding an independent estimate of the black hole mass and shock geometry.
Second, utilizing \textit{Insight-HXMT}, 
we extend the energy-dependent analysis of the LFQPO up to $\sim 100$ keV, which is significantly higher than prior studies to probe the modulation of the innermost Comptonizing 
corona.
Finally, we employ the well-established dynamic cospectral analysis on \textit{NuSTAR} data, allowing us to robustly track the rapid intra-observational evolution of the 
QPO frequency while completely bypassing the severe dead-time artifacts that typically plague high-count-rate observations.

The paper is structured as follows.
In \S1, we give a short introduction to the black hole properties as well as of the source.
In \S2, we describe the data reduction and analysis 
in detail. In \S3, we report the results of our analysis.
In \S4, we discuss the possible physical origin of our result.
Finally, in \S5, we summarize our result and draw notable 
conclusions.
\section{Observation, Data Selection, Reduction, and Analysis}

This source has been extensively studied by various X-ray satellites since its discovery.
For our analysis, we use data from the NICER, NuSTAR, and China's first X-ray satellite 
Insight-HXMT.
We discuss the data selection, reduction, and analysis in the subsequent subsections.
\subsection{Data Selection}

Several NICER observations available on a daily basis. NuSTAR also monitored the source on four separate days.
Insight-HXMT also monitored the source nearly on a daily basis.
The source had already been studied on a daily basis before.
Thus, we mainly focused our analysis on the simultaneous NICER-NuSTAR observations which were available on 2021 
May 05 (MJD 59339) and 2021 June 17 (MJD 59382).
HXMT data was available on MJD 59339 and it has two exposures.
We have taken it to study mainly the timing properties and also 
analyzed for spectral studies.
The selected data are listed in Table 1.

\begin{table}[h!]
\scriptsize
 \addtolength{\tabcolsep}{-1.0pt}
 \centering
\caption{List of Data used.
Column 1 lists the name of the satellite data that we have used for this work.
$^{[1]}$ Column 2 lists the observation IDs of those satellites, 
	 used for this work.
$^{[2]}$ Column 3 lists the time of those observation IDs in column 2. $^{[3]}$ Column 4 lists the corresponding MJDs of those observation IDs.
$^{[4]}$ 
	 Column 5 lists the exposures of those observation IDs.}
 \label{tab:table1}
 \begin{tabular}{ccccc}
 \hline
 Satellite      &   Obs.
Id.$^{[1]}$  &  UT$^{[2]}$    &  MJD$^{[3]}$   &  Exp.
(s)$^{[4]}$ \\
    (1)         &         (2)         &     (3)        &     (4)        &       (5)         \\
\hline
   NICER        &     4202130104      &   2021-05-05 
  &     59339      &       4039        \\ 
	        &     4675020124      &   2021-06-17   &     59382      &       2876        \\
\hline                    
                                                          
   NuSTAR       &     90702316002     &   2021-05-05   &     59339      &      
26554        \\
	        &     90702318003     &   2021-06-17   &     59382      &      15584        \\
\hline                                           
                                   
 Insight-HXMT   &     P0304014003     &   2021-05-05   &     59339      &      23325        \\
\hline
 \end{tabular}

\vspace{0.2cm}
\end{table}

\subsection{Data Reduction}

For this work, we have made use of the NICER, NuSTAR, and Insight-HXMT data.
Also, for all the data reduction and analysis processes, we have used the {\fontfamily{pcr}\selectfont HEASoft 
(version 6.32)} software.
We discuss our followed data reduction process below.

\subsubsection{NICER}

We use data from the X-ray timing instrument (XTI) on board the Neutron Star Interior Composition Array (NICER) at the International Space Station (ISS).
The XTI has a temporal resolution 
of $100 ns$ and energy resolution of $\sim$ 85 eV at 1 keV.
It covers an energy range of $0.2-10 keV$. For the whole data reduction process we have made use of the calibration file (version 
20221001; package {\fontfamily{pcr}\selectfont goodfiles\_nicer\_xti\_20221001.tar.gz}).
First, we ran the {\fontfamily{pcr}\selectfont nicerl2} task to generate cleaned event files from 
the raw ones.
Then using {\fontfamily{pcr}\selectfont nicerl3} command, we generated analyzable spectrum and light curve files in desired energy bands.
This process now incorporates the 
background corrections for the updated {\fontfamily{pcr}\selectfont CALDB}.
Finally, we use the {\fontfamily{pcr}\selectfont grppha} task to group the data for a minimum of 20 counts per 
spectral bin.
\subsubsection{NuSTAR}

We use both the focal plane module (FPM) A and B (FPMA and FPMB) instrument data of the NuSTAR satellite.
For our reduction purpose, we use the {\fontfamily{pcr}\selectfont NuSTARDAS
(version 2.1.0)} software and caldb files (version 20230705; package {\fontfamily{pcr}\selectfont goodfiles\_nustar\_fpm\_20230705.tar.gz}).
We first run the {\fontfamily{pcr}\selectfont
nupipeline} command to produce cleaned event files from the uncleaned raw data.
Then, using the {\fontfamily{pcr}\selectfont XSELECT} task, we choose source and background region files
by loading the cleaned event files in {\fontfamily{pcr}\selectfont ds9}.
As there was no evidence of pile-up, we used a circular region of 80 arcsec for both the source and background
and saved them.
Then, using those region files, we run the {\fontfamily{pcr}\selectfont nuproducts} command to produce science analyzable spectrum and light curve files.
We finally use
the {\fontfamily{pcr}\selectfont grppha} task to group the data with a minimum of 30 counts per spectral bin.
\subsubsection{Insight-HXMT}

We use the \href{http://hxmten.ihep.ac.cn/software.jhtml}{HXMTDAS (v2.05)}\footnote{http://hxmt.org/index.php/usersp/dataan} software for all our data reduction process.
After downloading
the level-1 data from the \href{http://archive.hxmt.cn/proposal}{HXMT data archive}\footnote{http://archive.hxmt.cn/proposal}, we first run the {\fontfamily{pcr}\selectfont hpipeline}
command to produce science-analyzable files.
This command runs several automatic subcommands to produce spectrum and light curve files from raw data files.
We provided suitable
conditions for the command to run properly with good time intervals.
For example, we set geomagnetic cutoff rigidity $> 8 GeV$, elevation angle $> 10^\circ$, pointing offset angle $< 
0.04^\circ$, and $> 600s$ away from the South Atlantic Anomaly (SAA).
Using the pipeline command, we have also produced light curves in various energy bands, as required for our
analysis purpose.
We finally group the spectrum using the {\fontfamily{pcr}\selectfont grppha} task with a minimum of 30 counts per bin.
\subsection{Data Analysis}

We have performed both the timing and spectral analysis of the 2021 outburst of the BHC MAXI J1803-298.
The daily average light curve has been downloaded from the 
\href{https://maxi.riken.jp/star\_data/J1803-298/J1803-298.html}{MAXI/GSC}\footnote{https://maxi.riken.jp/star\_data/J1803-298/J1803-298.html} website.
This work is not focused 
on the spectral classification of this outburst.
We are focused on two specific observations where NICER-NuSTAR simultaneous data were present.
In that regard, we used data from only two 
dates, as listed in Table 1. Insight-HXMT data was also available on 2021 May 5, and it is mainly used for timing analysis.
For timing analysis, we used mainly the NuSTAR and Insight-HXMT data. First, we discuss the NuSTAR data.
We first produced $0.01 sec$ time-binned light curves using the 
{\fontfamily{pcr}\selectfont nuproducts} command.
Then, using those light curves, we performed a fast Fourier transform (FFT) to produce power density spectrum (PDS) using the 
{\fontfamily{pcr}\selectfont powspec} command of the {\fontfamily{pcr}\selectfont XRONOS} package of the {\fontfamily{pcr}\selectfont HEASoft} software.
The data were subdivided into 
several intervals, containing 8192 newbins/interval.
For each interval, a PDS is generated and then all the PDS were averaged to produce a resultant PDS.
We choose the normalization 
such that the average PDS power will be in the $rms^2/Hz$ unit and the square root of the integral will give the $rms$ fractional variability.
White noise is subtracted and we use 
$-1.02$ as the geometrical rebinning factor.
Using the PDS, we searched for low-frequency QPOs (LFQPOs). We also make the cos-spectrum with lightcurves from the FPMA and FPMB modules 
of NuSTAR, and dynamic PDS using the \href{https://github.com/StingraySoftware/stingray}{StingRay}\footnote{https://github.com/StingraySoftware/stingray} software package.
We discuss 
this in the result section. Along with these, as we mentioned before, we selected 7 GTIs from one of the NuSTAR observation IDs on 2021 May 5. Using those GTIs also, we produced PDS in 
the same way, as mentioned above.
We searched for LFQPOs in these GTIs. In all these PDSs, we fit the QPO nature using the Lorentzian model.
For the HXMT data, we produced light curves 
of the three modules, LE, ME, and HE.
Using those light curves also, we produced PDSs and searched for QPOs.
If a QPO feature is present, we fitted with a Lorentzian model to extract 
QPO properties like QPO frequency ($\nu_{qpo}$), full width at half maximum (FWHM), and QPO normalization.
Using the HXMT HE light curve, we produced light curves in six different 
energy bands (27-35, 35-48, 48-67, 67-100, 100-150, and 150-200 ~keV) and produced PDS to determine the energy dependence of QPOs.
We discuss these in detail in the result section.

For spectral analysis, we have mainly used simultaneous NICER-NuSTAR data on 2021 May 5 (MJD 59339) and 2021 June 17 (MJD 59382).
Using NICER-NuSTAR, we have spectrally fitted the data
in the 1-70 ~keV energy range.
For spectral analysis, we have used various combinations of models, to achieve the best fit.
However, we checked the spectral fitting for NICER and 
NuSTAR data separately in the 1-10 and 3-70 ~keV energy bands respectively.
First, we fitted the NICER data in the 0.5-10 ~keV energy band.
Due to the presence of the instrumental 
residuals below 1 ~keV, we have refitted and taken above 1 ~keV.
For all the data, {\fontfamily{pcr}\selectfont tbabs} model is used in order to account for the interstellar absorption.
For the data on 2021 May 5, for only NICER data, good fits were achieved using {\fontfamily{pcr}\selectfont tbabs*(diskbb + power-law + gaussian)}, and {\fontfamily{pcr}\selectfont 
tbabs*(diskbb + nthcomp + gaussian)} models, whereas, for data on 2021 June 17, NICER data was fitted using {\fontfamily{pcr}\selectfont tbabs*(diskbb + power-law)} and {\fontfamily{pcr}\selectfont 
tbabs*(diskbb + nthcomp)} models.
For NuSTAR data, on 2021 May 5, a statistically acceptable fit was achieved using a combination of several models.
Those are: {\fontfamily{pcr}\selectfont 
tbabs*(nthcomp + gaussian)}, {\fontfamily{pcr}\selectfont tbabs*(broken power-law + gaussian)}, and {\fontfamily{pcr}\selectfont tbabs*(relxill + gaussian)} models, whereas, for data 
on 2021 June 17, this was achieved using  {\fontfamily{pcr}\selectfont tbabs*(diskbb + broken power-law)} model.
The broadband data, on 2021 May 5, is fitted for an acceptable $\chi^2_{red}$ 
using {\fontfamily{pcr}\selectfont constant*tbabs(diskbb + broken power-law + gaussian)} and {\fontfamily{pcr}\selectfont constant*tbabs(nthcomp + relxill + gaussian)} models, whereas, 
for data on 2021 June 17, it was achieved using {\fontfamily{pcr}\selectfont constant*tbabs(diskbb + broken power-law)} and {\fontfamily{pcr}\selectfont constant*tbabs(diskbb + relxill)} 
models.
Besides these spectral analysis, we also selected and analyzed data of different good time intervals (GTIs) of the NuSTAR observation.
On 2021 May 5, NuSTAR data show some absorption 
dips in nature.
Using the cleaned event file in the {\fontfamily{pcr}\selectfont XSELECT}, we selected 7 GTIs and generated light curves and spectrum files for each of these 7 GTIs.
Using 
those spectrum files, we also fitted NuSTAR data using the combination of {\fontfamily{pcr}\selectfont tbabs*(diskbb + gaussian + broken power-law} and {\fontfamily{pcr}\selectfont 
tbabs*(gaussian + broken power-law)} models. We also checked the spectral fitting using the physical two-component advective flow (TCAF) model (Chakrabarti \& Titarchuk 1995; Debnath et al. 2014; Mondal et al. 2014; Bhattacharya et al. 2024) on 2021 May 
05 data. We found that the best fit was achieved using the combination of {\fontfamily{pcr}\selectfont tbabs*(TCAF + gaussian + relxill)} model. The best fit will be discussed in the result 
section. Since this absorption dip nature was not observed in the observation on 2021 June 17, we did not make any GTI 
selection for it. The resulting best-fit parameters, physical 
interpretations, and spectral state classifications derived from these model combinations are presented and discussed in Section 3.3.

\section{Results}

We have performed both timing and spectral analysis of this source using data from two epochs 2021 May 05 (MJD 59339) and 2021 June 17 (MJD 59382) using NICER, NuSTAR, and Insight-HXMT data.
We would like to show the results in this section.
We first show the outburst light curve variation along with its hardness ratio (HR) and hardness intensity diagram, then we show the
timing and spectral properties in the subsequent sections.
\subsection{Outburst Profile, Hardness Ratio (HR), and Hardness Intensity Diagram (HID)}

\begin{figure}[!h]
\centering
Evolution of LC, HR, and HID\par\medskip
\vbox{
\includegraphics[width=8.5truecm,angle=0]{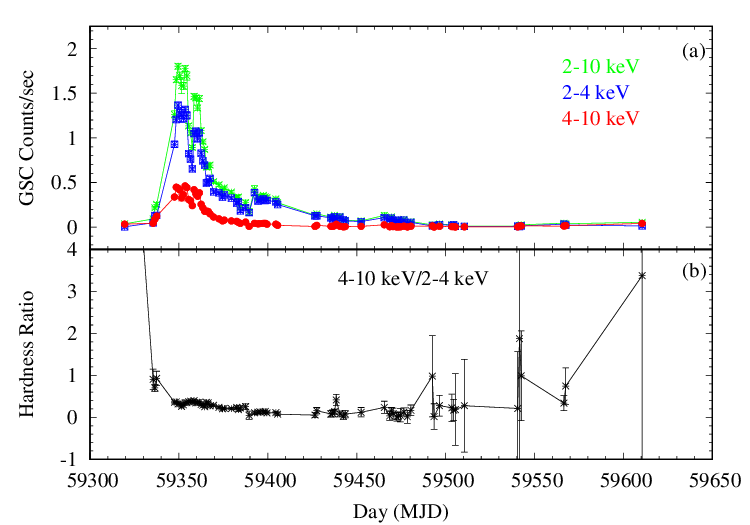}
\hskip 0.5cm
\includegraphics[width=8.5truecm,angle=0]{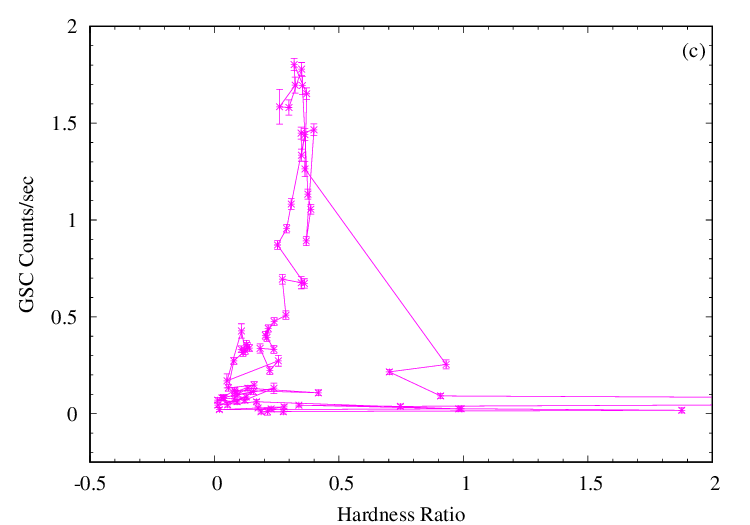}}
\caption{Variation of (a) MAXI/GSC count rates in 2-10, 2-4, 4-10 ~keV energy bands, (b) hardness ratio (HR; 4-10 ~keV/2-4 ~keV), and (c) hardness intensity diagram (HID) with time.}
\end{figure}

In Figure 1, we show the evolution of the light curve, HR, and HID during the entire duration of the outburst.
In panel (a), we show the MAXI/GSC count rates in the 2-10 (green), 2-4 (blue), 
and 4-10 ~keV (red) energy bands.
In panel (b) the hardness is shown using the ratio 4-10 ~keV count rate to 2-4 ~keV count rates.
In panel (c), we show the variation of the 2-10 ~keV 
count rate with the HR.
Looking at the light curve, it can be noticed that the flux rose very fast to its peak after the onset of the outburst.
Then the total flux decayed quite quickly for the first few days after the
peak.
However, after that, it showed slow decay and reached the quiescence after a comparatively long time.
The HR was slightly high at the start of the outburst, then it decreased to a lower 
value, and remained low for the majority of the outburst.
At the end of the outburst, it again rose to higher values.
The HID shows a loop-like behavior, that is common for stellar-mass
black holes.
HR roughly approximates the spectral nature of an outburst. When HR was higher, counts were lower, and as HR decreased the count rate was increased.
This is common behavior for SBHs.
Judging by the variation of the light curve, we could say that the outburst followed a fast rise slow decay (FRSD) profile and from the variation of HR, and HID, we designate that the source went
through all the defined spectral states of a BH.
Studying spectral evolution is not the scope of this paper.
Thus we have not designated any transition dates between different spectral states.
\subsection{Timing Properties}

\begin{figure}[!h]
\centering
NuSTAR light curve\par\medskip
 \vbox{
 \includegraphics[width=6.5truecm,angle=270]{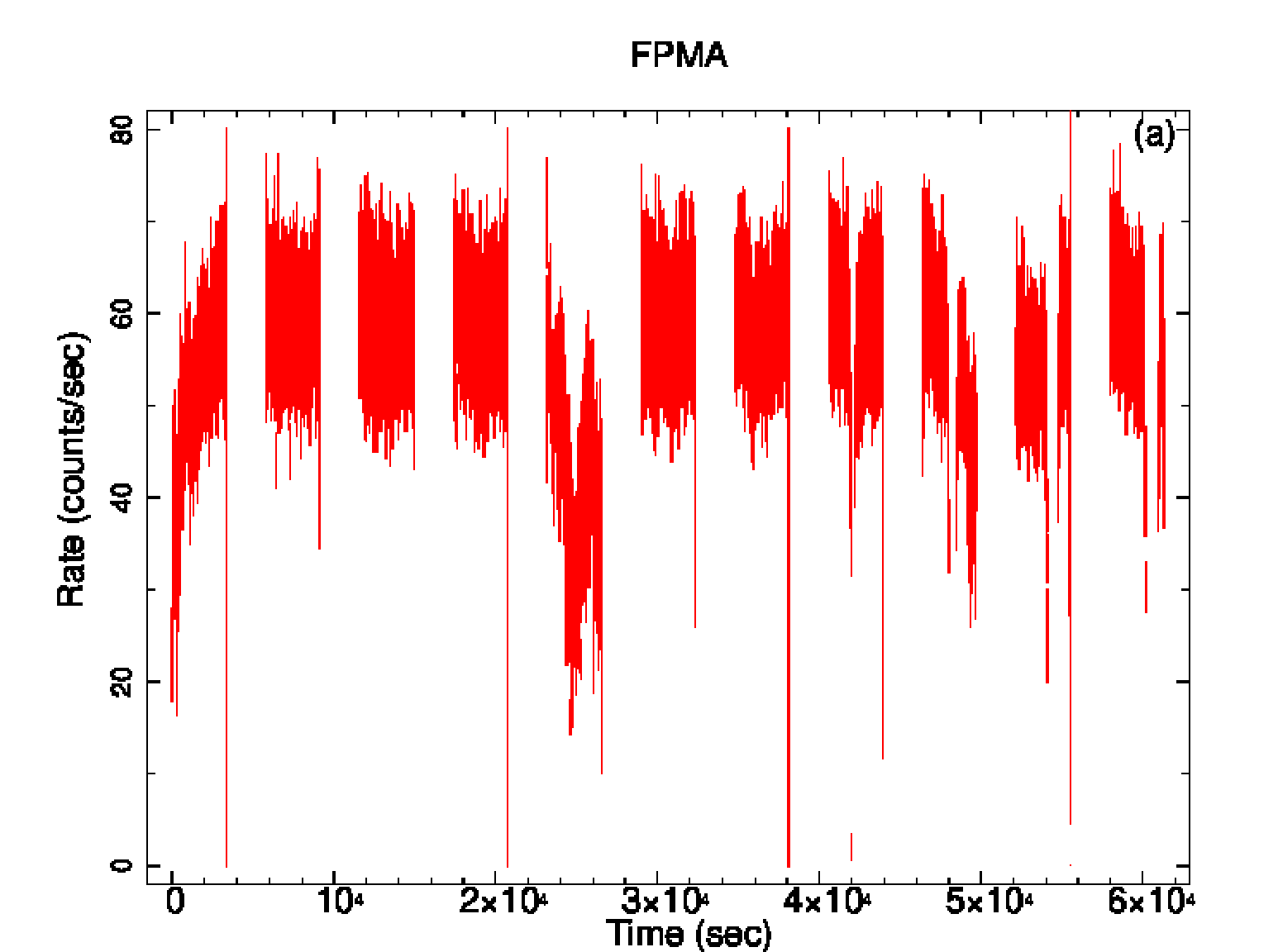}
 \hskip 0.5cm
 \includegraphics[width=6.5truecm,angle=270]{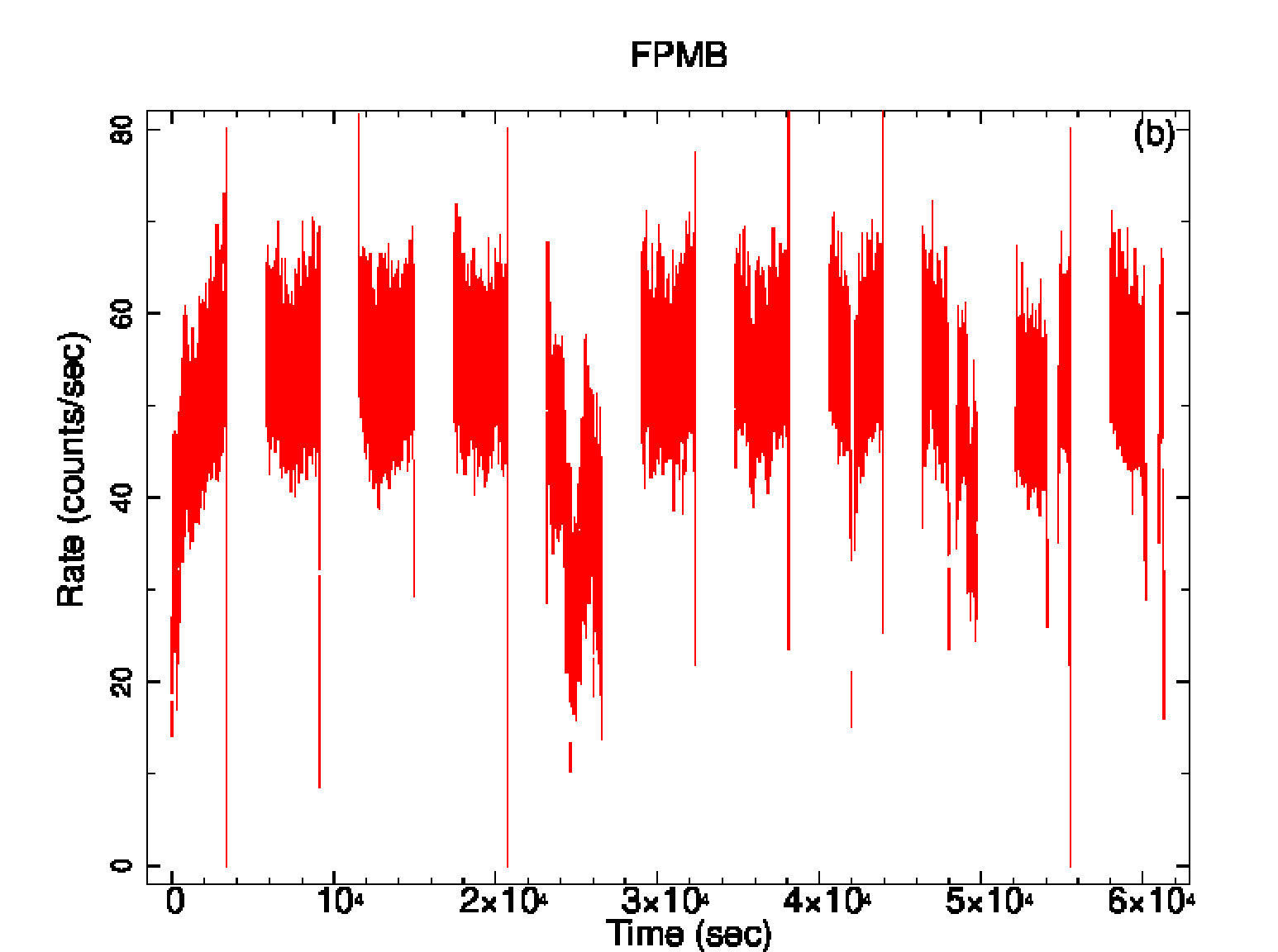}}
 \caption{$1 ~sec$ time-binned light curve data for (a) FPMA, (b) FPMB from NuSTAR satellite for the obs.
ID. 90702316002. It shows a dip nature in both modules.}
\end{figure}

To explore the timing properties, we primarily studied the quasi periodic oscillations, by producing power density spectrum (PDS) from 0.01~s time-binned lightcurves.
However, in timing analysis, 
we implemented several different approaches. From the NuSTAR observation ID 90702316002, we first produced $1$ s binned light curves for both FPMA and FPMB modules.
We found that there are data 
gaps in both the light curves, as can be noticed from Figure 2. Along with that, we noticed that in some segments, the count rate was lower than in other segments.
This is due to the fact that 
this source is in a 7-hour period binary system (Homan et al. 2021; Jana et al. 2021) and the inclination is high.
We then produced the white noise non subtracted Leahy normalized 
PDS from the 0.001 s time-binned lightcurves of both the FPMA and FPMB modules which are shown in Figure 3. From the figure, we can see that although a QPO-like peak is present in the PDSs at $\sim$ 
0.4 Hz, the power drops below 2 above $\sim1$ Hz and again increases above $\sim$ 100 Hz.
This is a result of the distortion of the white noise level arising out of the long ($\sim$2.5 ms) and variable 
instrumental dead time of the NuSTAR satellite.
Such effects are much more pronounced in Galactic X-ray binary sources that typically have count rates $\gtrsim$ 100 cts s$^{-1}$ (Bachetti et al. 
2015).
\begin{figure}
\centering
PDS generated from NuSTAR data\par\medskip
\vbox{
\includegraphics[width=9.0truecm,keepaspectratio=true]{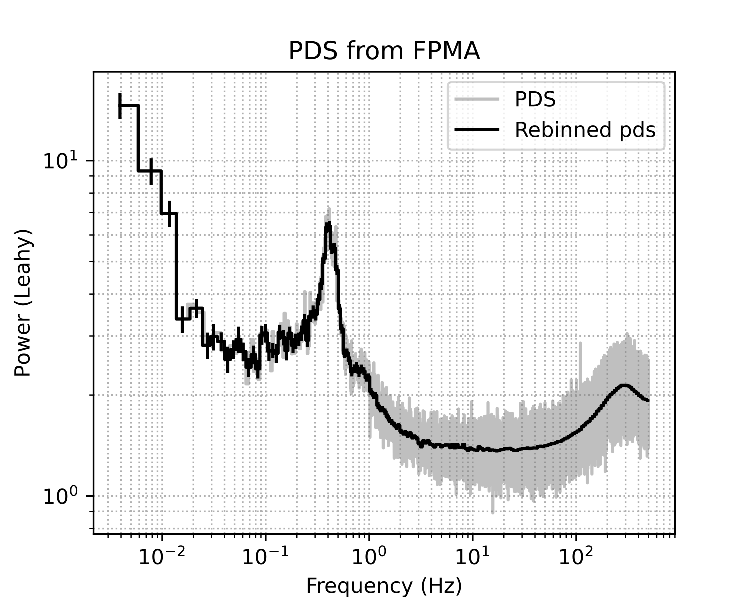}
\hskip 0.5cm
\includegraphics[width=9.0truecm,keepaspectratio=true]{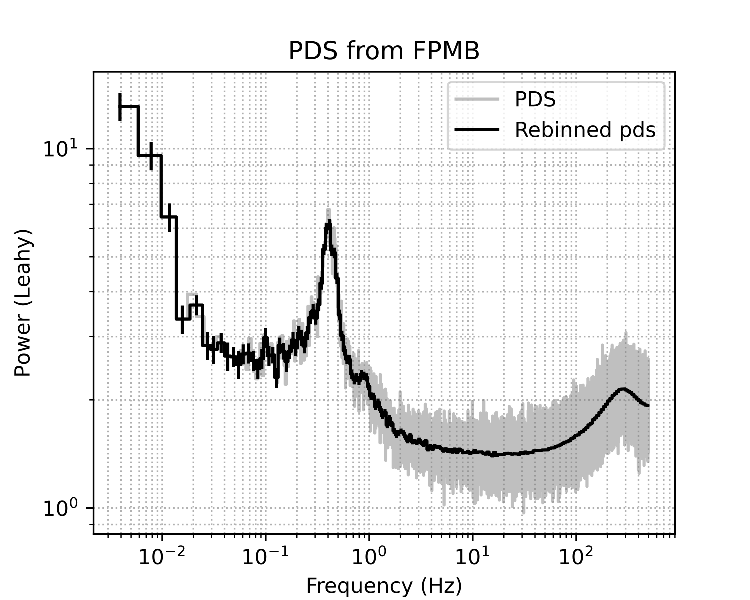}}
\caption{White noise non-subtracted power density spectra using 0.001 s time-binned light curve data of FPMA (top), and FPMB (bottom) for the obs.
ID. 90702316002.}
\end{figure}

\begin{figure}[!h]
  \centering
    \includegraphics[width=9.0truecm]{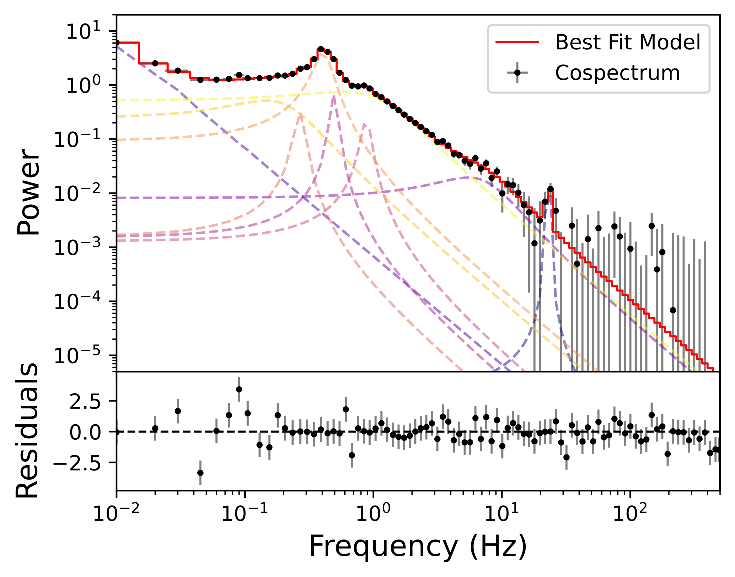}
    \caption{Cospectrum fitted using multiple Lorentzian models for the obs.
ID. 90702316002.}
\end{figure}

To mitigate the effects of dead time, following Bachetti et al.
(2015), we constructed the \textit{cospectrum} which is the real part of the complex cross-spectrum between the lightcurves from 
FPMA and FPMB.
This \textit{cospectrum} effectively provides the same information as the white noise subtracted PDS.
We fitted this \textit{cospectrum} with multiple Lorentzian profiles that is 
shown in Figure 4. From the figure, we see that the high frequency bump is absent in the \textit{cospectrum} and there is a fundamental QPO present at $\sim 0.4$~Hz.
\begin{figure}[!h]
\vspace{0.6cm}
  \centering
    \includegraphics[width=8cm]{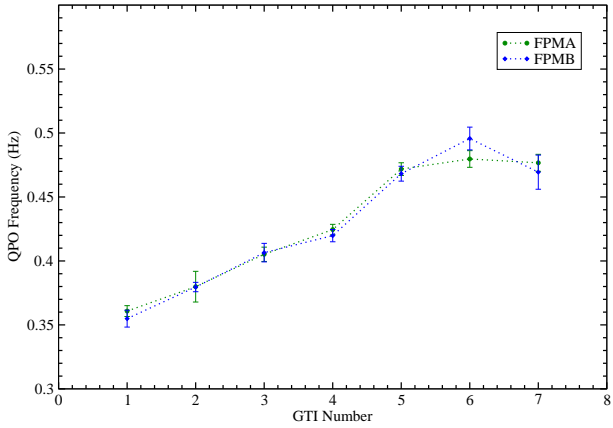}
    \caption{Variation of QPO frequency from 7 different GTIs using both the FPMA and FPMB modules for the obs.
ID. 90702316002.}
\end{figure}

We also cut 7 good time intervals (GTIs) from the NuSTAR light curve in both FPMA and FPMB modules, that is shown in Figure 2. Using these GTIs, we produced $0.01$~s time-binned light curves to 
study the evolution of QPO frequency within this day.
We found that QPO frequency ($\nu_{qpo}$) evolved within this day.
This can be observed in Figure 5. $\nu_{qpo}$ in the range of 0.35-0.47
Hz and 0.35-0.49~Hz for FPMA and FPMB, respectively.
Both of these modules showed very similar evolution, as can be seen from this figure.
\begin{figure}[!h]
  \centering
    \includegraphics[width=10cm]{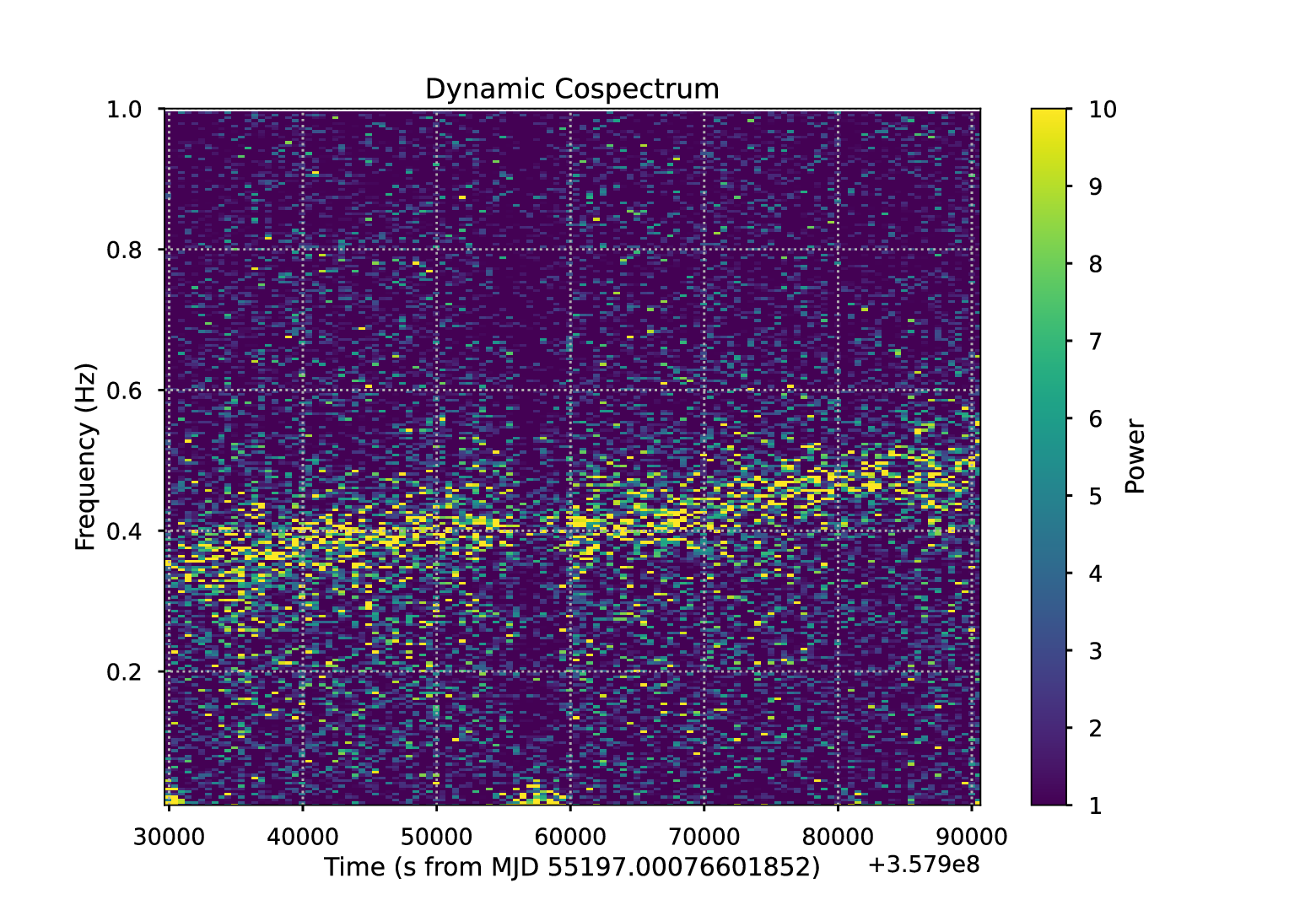}
    \caption{Dynamic cospectrum using NuSTAR data for the obs. ID.
90702316002.}
\end{figure}

To study the rapid variation of the QPO frequency within the observation duration, we also generated the dynamic \textit{cospectrum} from the light curves of FPMA and FPMB.
This is shown in Figure 6.
From the figure, it can be noticed that the QPO frequency evolved from $\sim 0.35$ Hz to $\sim 0.50$ Hz within the period of observation.
\begin{figure*}
\centering
PDS generated from HXMT data\par\medskip
\vbox{
\includegraphics[width=4.1truecm,angle=270]{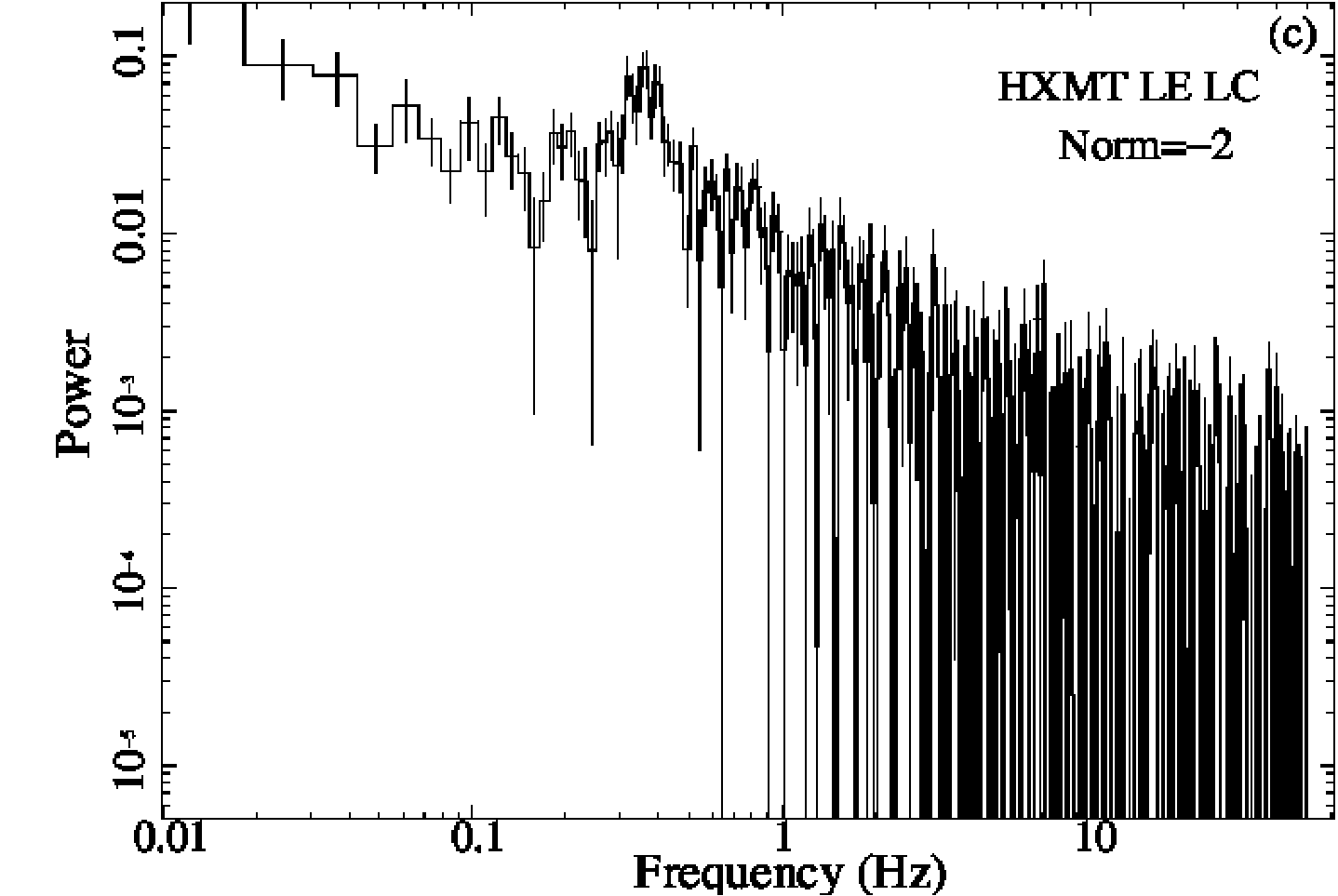}
\includegraphics[width=4.1truecm,angle=270]{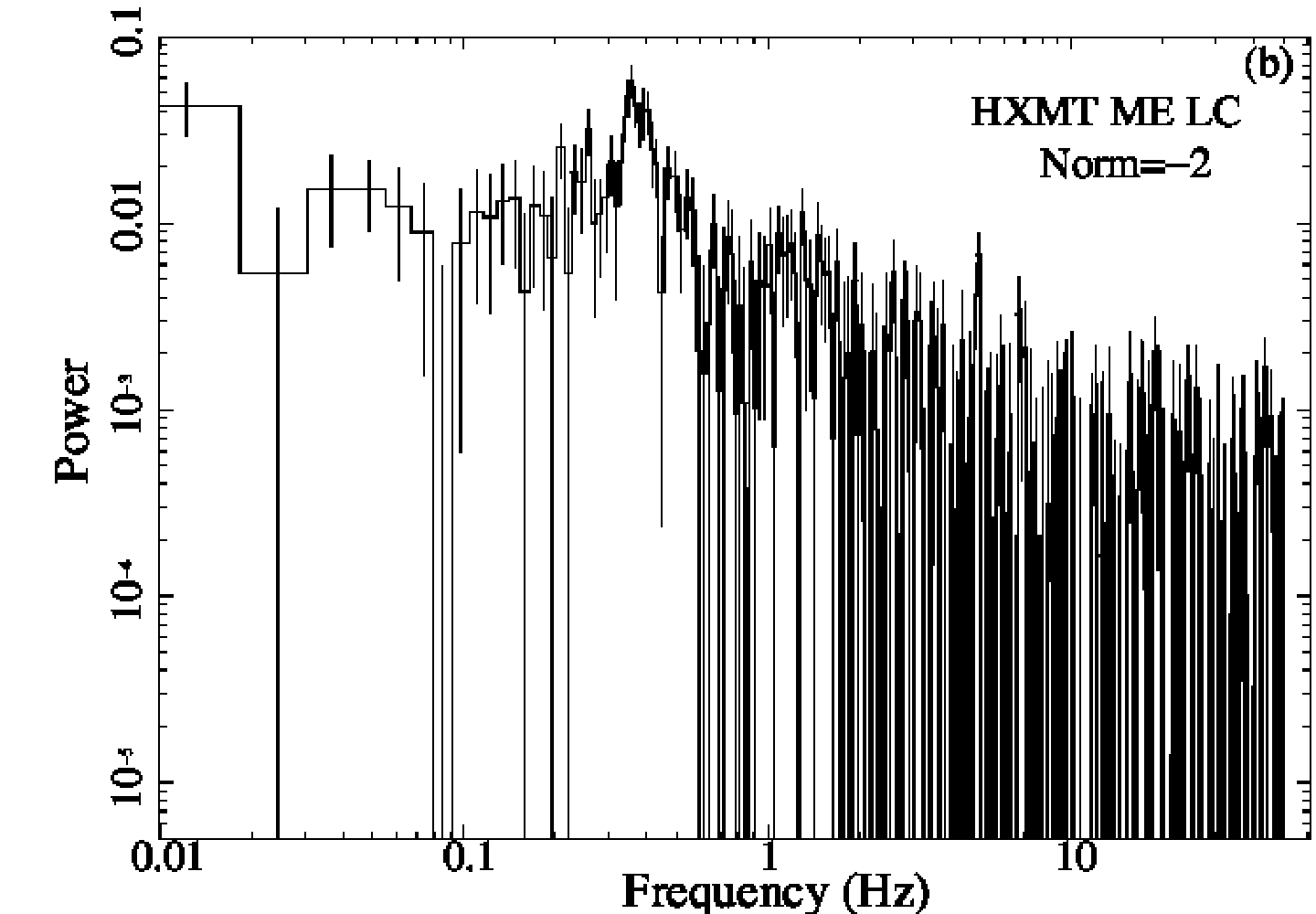}
\includegraphics[width=4.1truecm,angle=270]{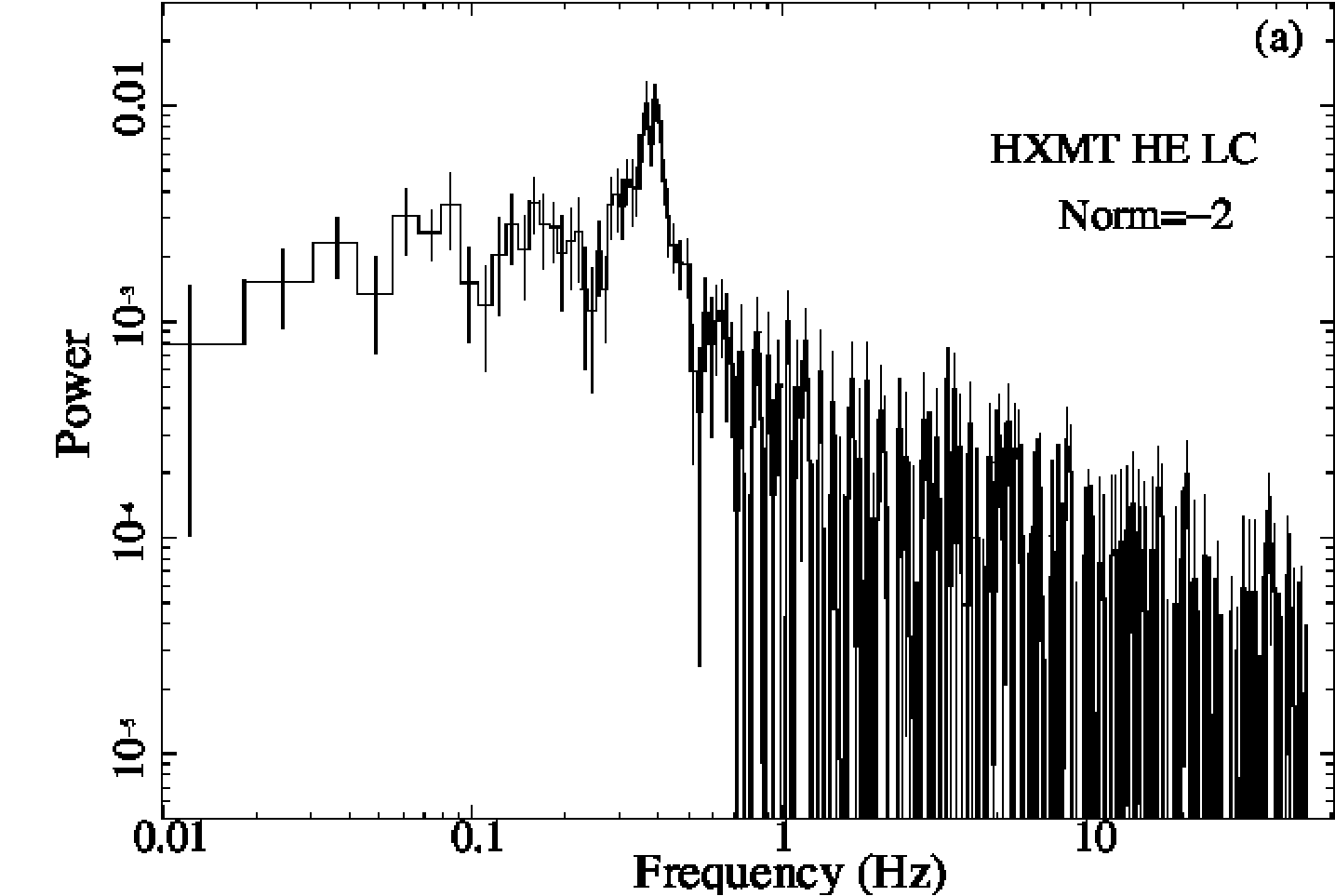}}
\caption{Power density spectra using $0.01 ~sec$ time-binned light curves for (a) LE, (b) ME, and (c) HE instruments of Insight-HXMT satellite.
This is for the obs. ID. P030401400301
         (exposure ID. P030401400301-20210505-01-01).}
\end{figure*}

\begin{figure*}
\centering
Energy Dependent PDS using HE data\par\medskip
\vbox{
\includegraphics[width=3.9truecm,angle=270]{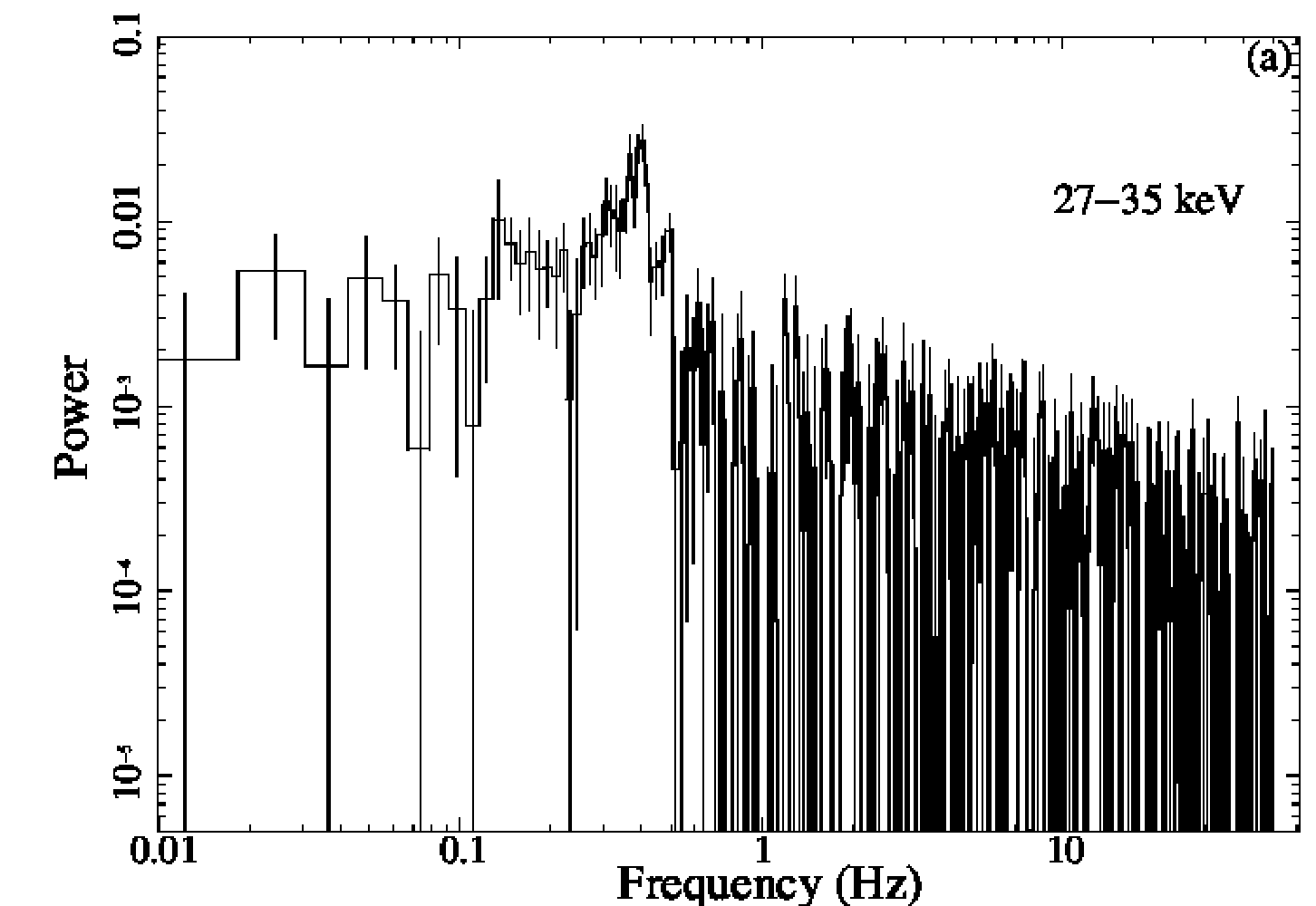}
\hskip 0.5cm
\includegraphics[width=3.9truecm,angle=270]{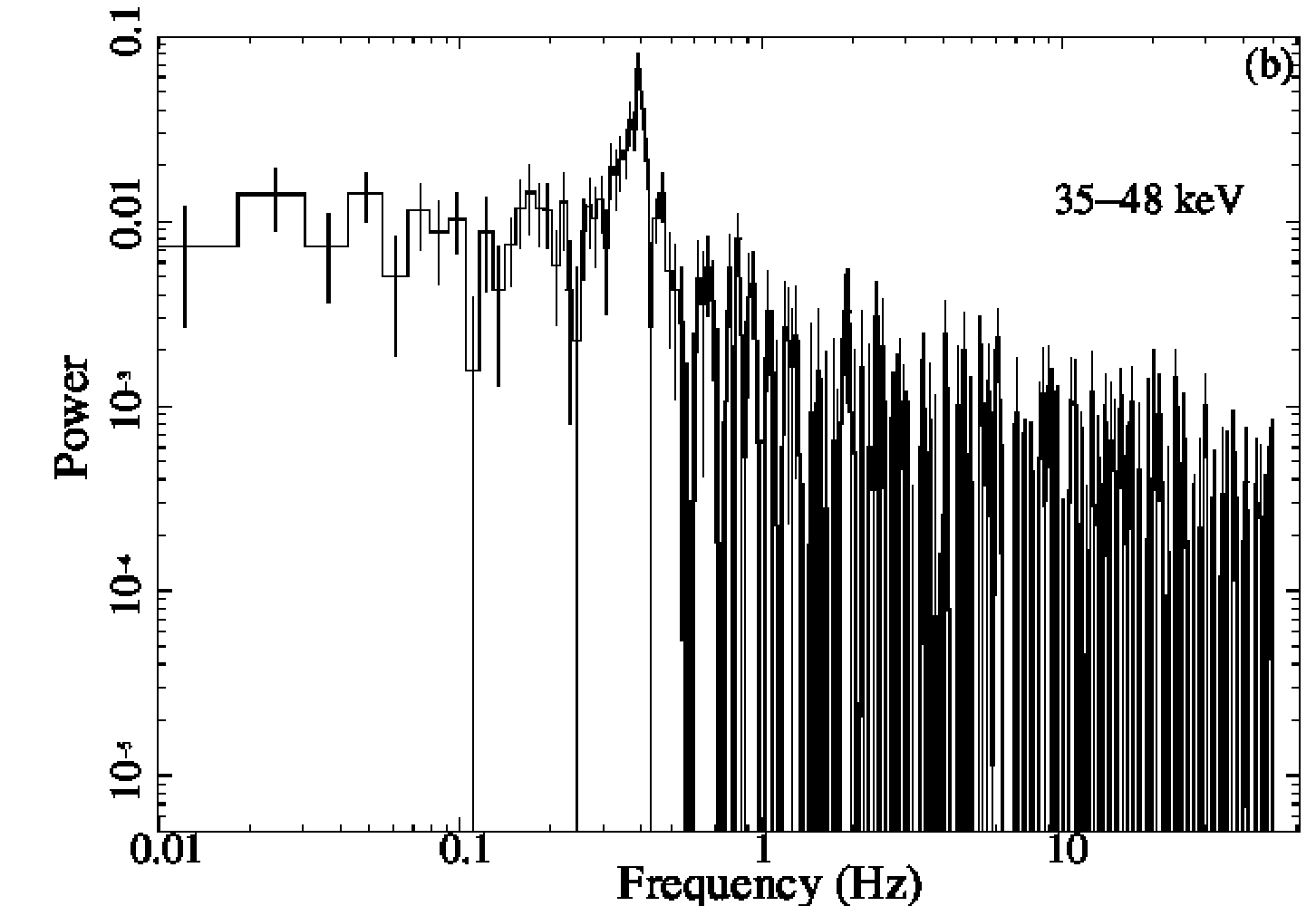}
\hskip 0.5cm
\includegraphics[width=3.9truecm,angle=270]{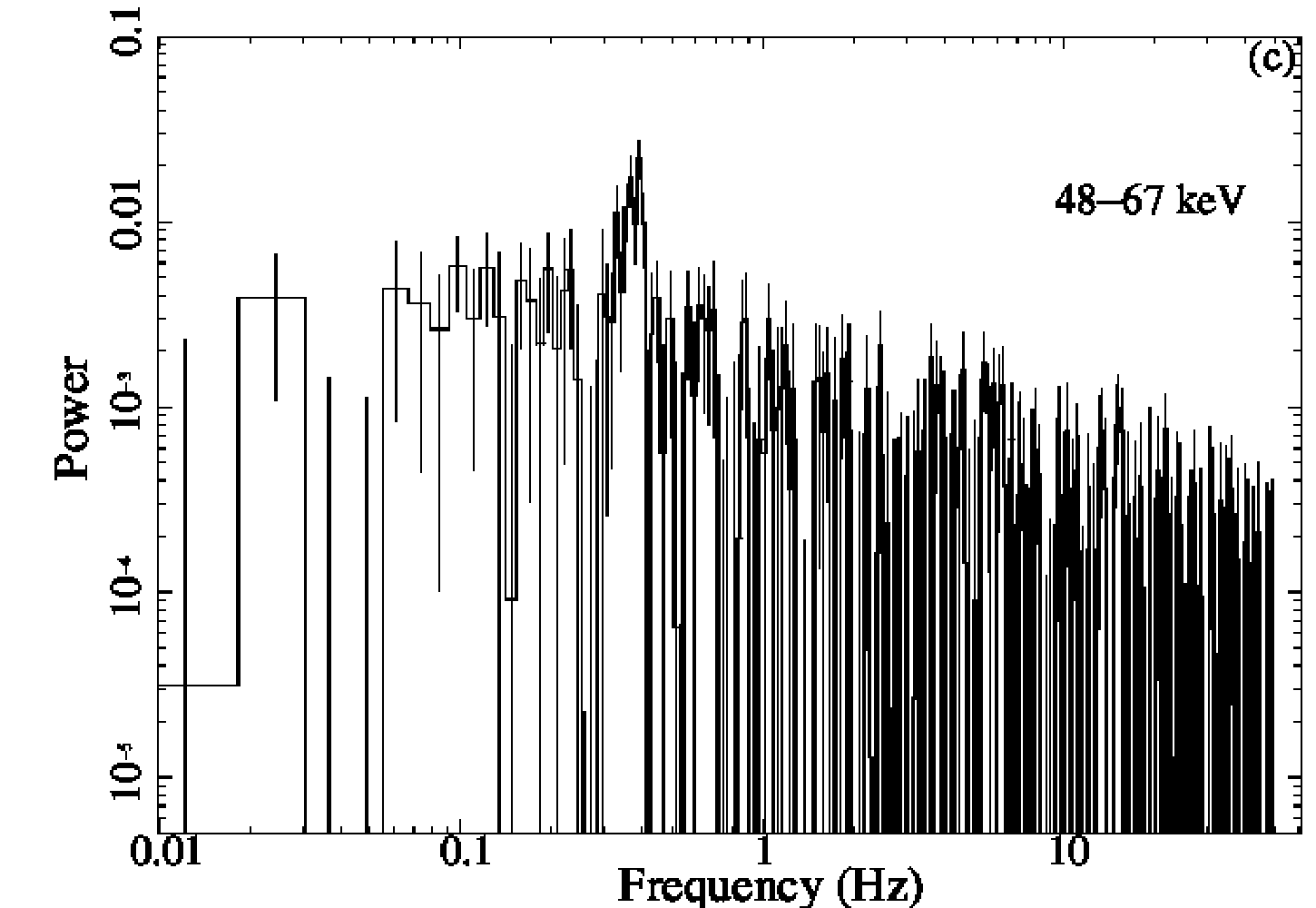}
\hskip 0.5cm
\vskip 0.5cm
\includegraphics[width=3.9truecm,angle=270]{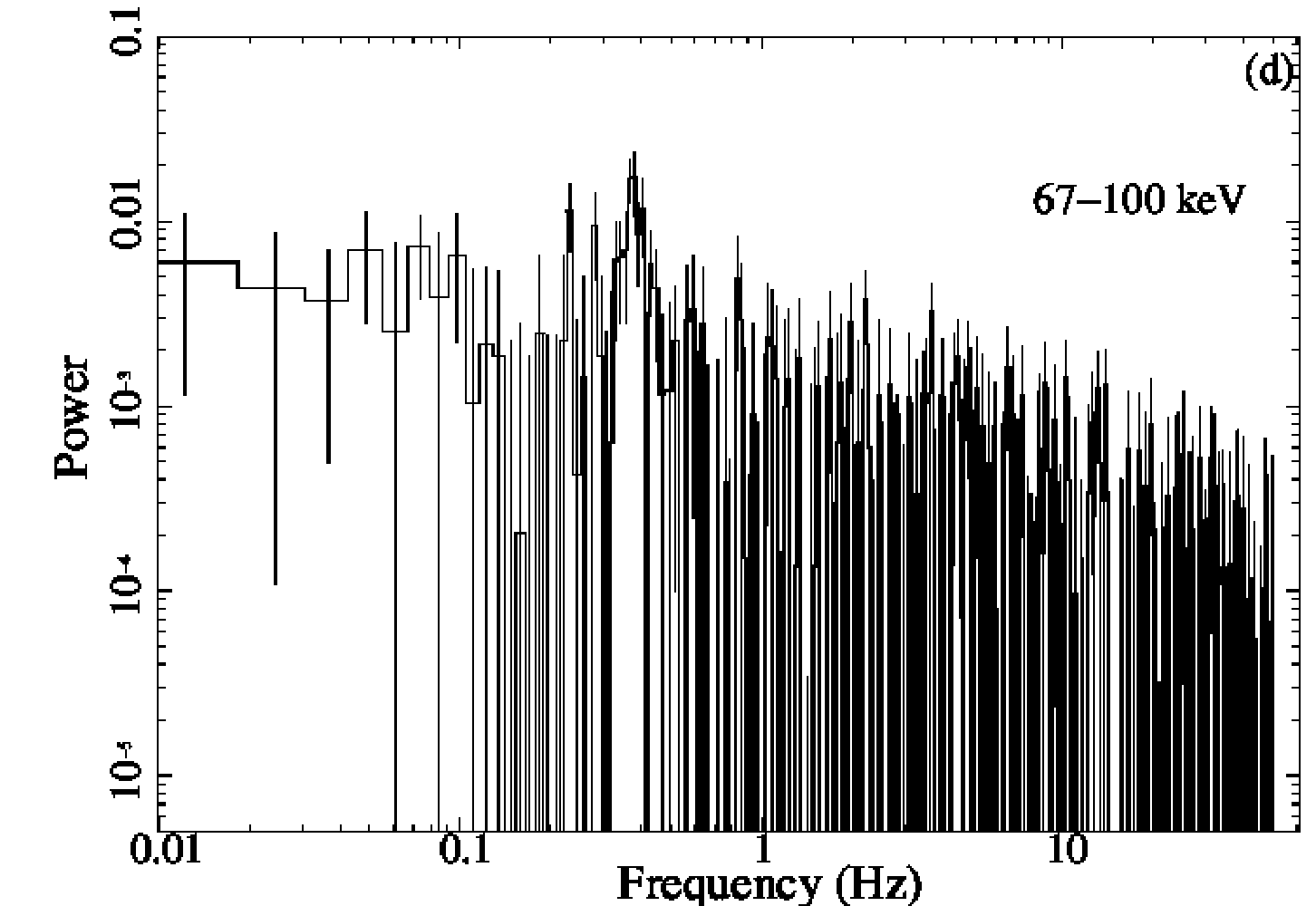}
\hskip 0.5cm
\includegraphics[width=3.9truecm,angle=270]{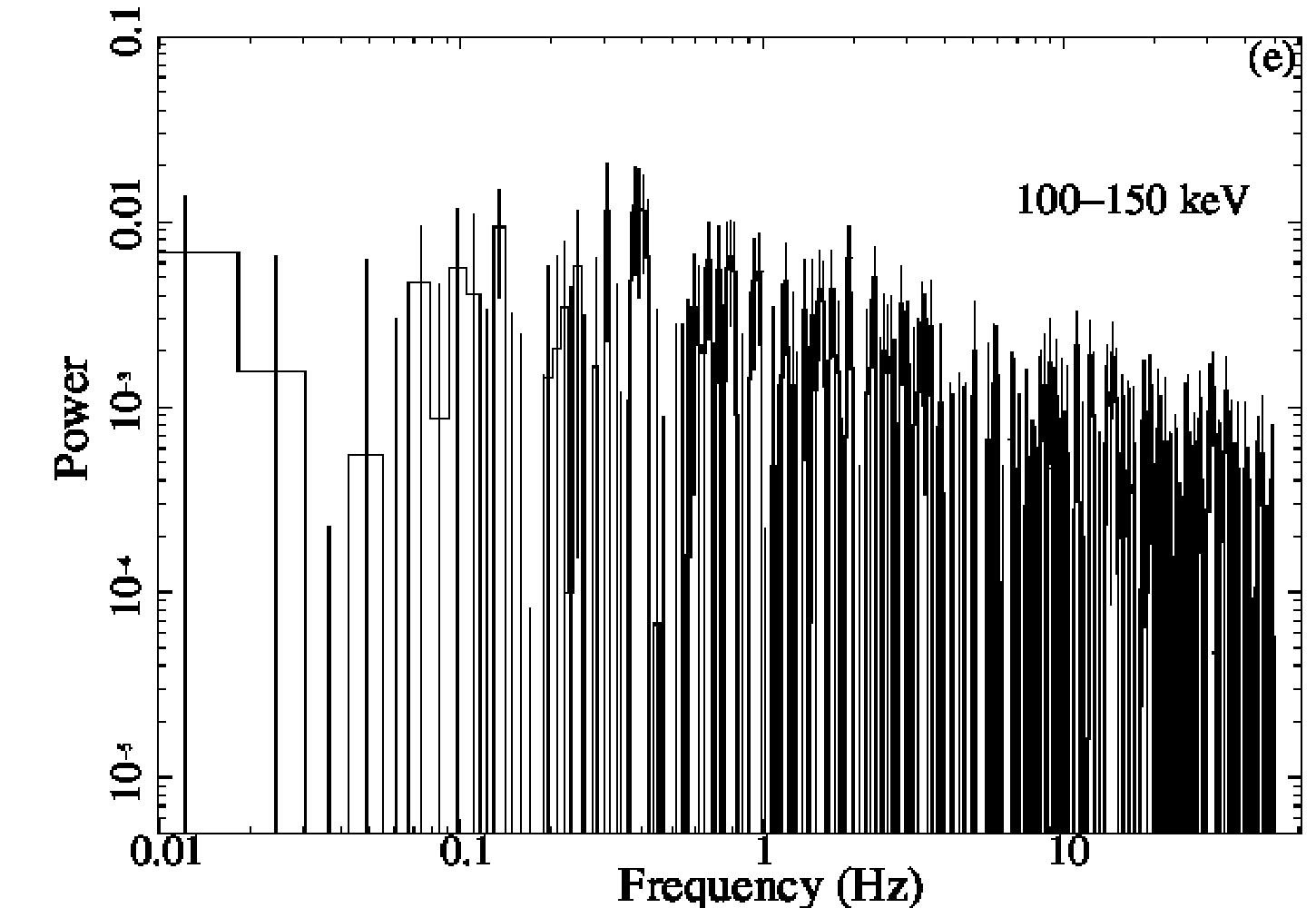}
\hskip 0.5cm
\includegraphics[width=3.9truecm,angle=270]{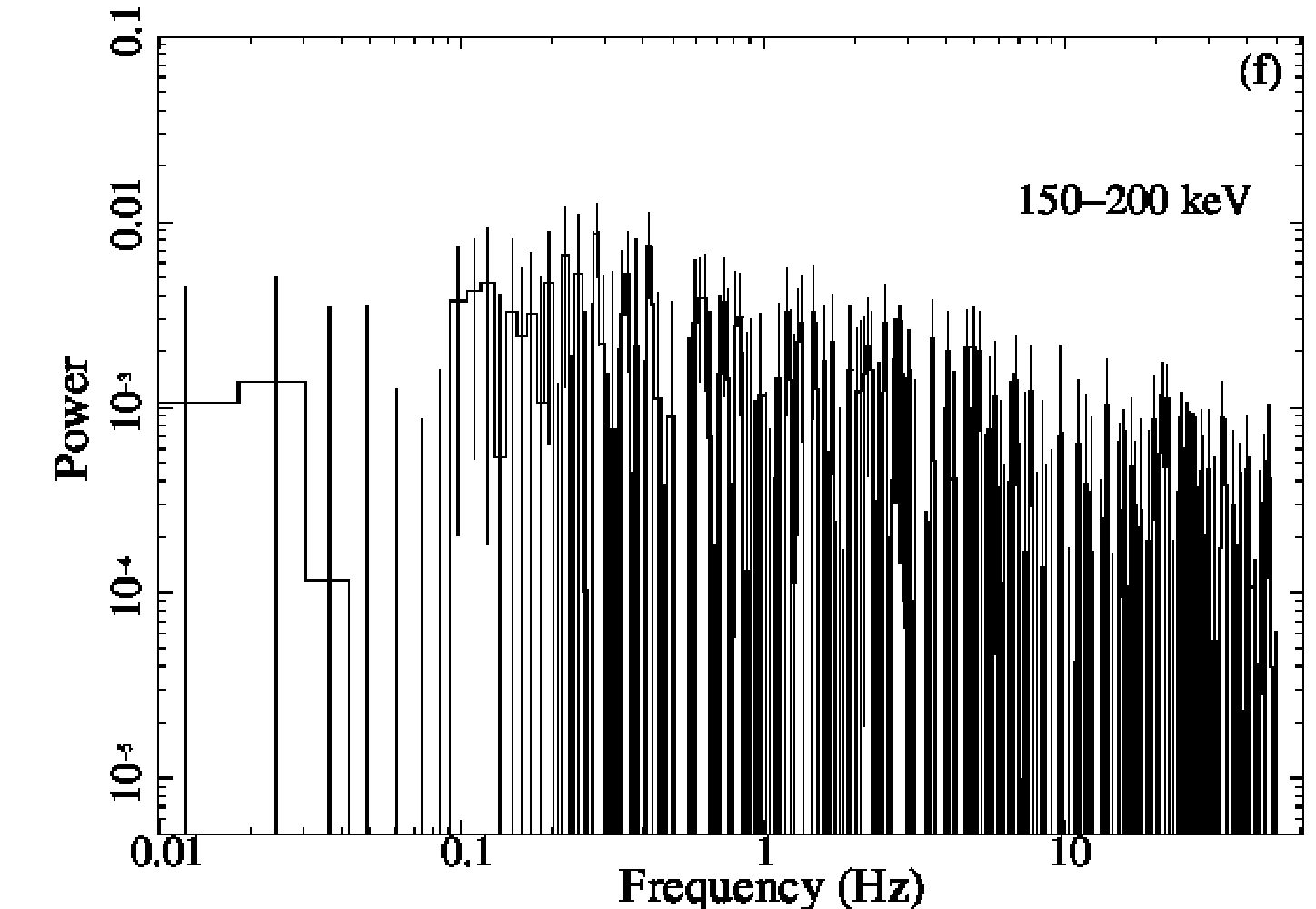}}
\caption{Power density spectra in (a) 27-35, (b) 35-48, (c) 48-67, (d) 67-100, (e) 100-150, and (f) 150-200 ~keV energy bands using $0.01 ~sec$ time-binned HE light curves 
         from Insight-HXMT.
This is for the obs. ID. P030401400301 (exposure ID. P030401400301-20210505-01-01).}
\end{figure*}

The change of QPO properties with different GTIs is listed in Table 2. For a comparison, we also produced PDS using the \textit{Insight}-HXMT data in LE (2-10~keV), ME (10-35~keV), and
HE (27-250~keV) bands by producing 0.01~s time-binned light curves first.
Unlike NuSTAR, this produces the full PDS up to 50~Hz which is the Nyquist frequency for a 0.01~s light curve.
This is shown in Figure 7. The QPO properties from \textit{Insight}-HXMT data is shown in Table 3. We notice that the QPO frequency shows slight variations in the three energy bands.
$Q$-value
varied between 2.5-3.8.  We show source and background count rates in columns 7 and 8. Using these values we estimated RMS for three bands using the formula from Bu et al.
(2015). These values 
are given in the column 9. We find that the RMS is lowest in the HE band, whereas it increased as we go to the lower energy bands.
\begin{table*}[!h]
	\vspace{0.5cm}
\scriptsize
 \addtolength{\tabcolsep}{-1.0pt}
 \centering
\caption{QPO Properties from different GTIs of NuSTAR data.
Column 1 is the numer of GTIs that we cut from the full NuSTAR light curve.
Columns 2-6 represent the QPO properties, extracted from 
	 the light curves of those GTIs for FPMA module.
Columns 7-11 represent the QPO properties, extracted from the light curves of those GTIs for FPMB module.}
 \label{tab:table2}
 \begin{tabular}{|c|ccccc|ccccc|}
 \hline
  GTI   &                                     \multicolumn{5}{|c|}{FPMA}                                       
  &                                         \multicolumn{5}{|c|}{FPMB}                                    \\
\hline                      
                                                                                         
 Number &     $\nu_{qpo}$   &  
       FWHM       &       Norm        &     $Q$-value      &      RMS (\%)        &      $\nu_{qpo}$    &      FWHM         &       Norm        &     $Q$-value 
     &      RMS (\%)       \\
  (1)   &         (2)       &        (3)        &       (4)         &       (5)          &        (6)    
        &         (7)         &       (8)         &       (9)         &       (10)        &       (11)          \\
\hline

   1    & $0.361 \pm 0.004$ & $0.221 \pm 0.018$ & $0.145 \pm 0.009$ & $1.633 \pm 0.1342$ & $22.437 \pm 1.148$   &   
$0.354 \pm 0.006$ & $0.231 \pm 0.014$ & $0.180 \pm 0.011$ & $1.532 \pm 0.096$ & $25.558 \pm 1.099$  \\         
   2    & $0.379 \pm 0.012$ & $0.187 \pm 0.017$ & $0.143 \pm 0.008$ & $2.026 \pm 0.1951$ & $20.496 \pm 1.093$   &   $0.379 \pm 0.003$ & $0.222 \pm 0.007$ & $0.159 \pm 0.011$ & $1.707 \pm 0.055$ & $23.548 \pm 0.894$  \\         
   3    & $0.405 \pm 
0.005$ & $0.133 \pm 0.027$ & $0.176 \pm 0.020$ & $3.045 \pm 0.6193$ & $19.176 \pm 2.229$   &   $0.406 \pm 0.007$ & $0.216 \pm 0.015$ & $0.170 \pm 0.011$ & $1.879 \pm 0.134$ & $24.018 \pm 1.139$  \\         
   4    & $0.424 \pm 0.004$ & $0.230 \pm 0.019$ & $0.126 \pm 0.007$ & $1.843 \pm 0.1532$ & $21.337 \pm 1.061$   &   $0.419 \pm 0.004$ & $0.252 \pm 0.023$ & $0.127 \pm 0.007$ & $1.662 \pm 0.152$ & $22.422 
\pm 1.195$  \\         
   5    & $0.471 \pm 0.004$ & $0.145 \pm 0.010$ & $0.170 \pm 0.007$ & $3.248 \pm 0.2257$ & $19.678 \pm 0.790$   &   $0.468 \pm 0.005$ & $0.187 \pm 0.012$ & $0.159 \pm 0.010$ & $2.502 \pm 0.162$ & $21.612 \pm 0.970$  \\         
   6    & $0.479 \pm 0.006$ & $0.153 \pm 0.028$ & $0.136 \pm 0.018$ & $3.130 \pm 0.5742$ & $18.080 \pm 2.041$ 
  &   $0.495 \pm 0.008$ & $0.279 \pm 0.040$ & $0.107 \pm 0.011$ & $1.774 \pm 0.255$ & $21.656 \pm 1.909$  \\         
   7    & $0.476 \pm 0.006$ & $0.138 \pm 0.025$ & $0.149 \pm 0.020$ & $3.449 \pm 0.6263$ & $17.973 \pm 2.025$   &   $0.469 \pm 0.013$ & $0.370 \pm 0.059$ & $0.100 \pm 0.009$ & $1.267 \pm 0.205$ & $24.109 \pm 2.206$  \\        
 \hline
 \end{tabular}
	\vspace{0.5cm}
\end{table*}

\begin{table*}[!h]
	\vspace{0.5cm}
\scriptsize
 \addtolength{\tabcolsep}{-1.0pt}
 \centering
	\caption{QPO Properties 
from Insight-HXMT light curve on May 2021 05 (Observation ID: P030401400).
Here $\sigma$ represents the significance of the QPO.} 
 \label{tab:table3}
 \begin{tabular}{ccccccccc}
 \hline
     Exposure   & Energy Band &      $\nu_{qpo}$    &         FWHM        &         Norm          & $Q$-value  &  Source Count  &  Background Count  &    RMS (\%)  \\  
        (1)  
    &     (2)     &           (3)       &          (4)        &         (5)           &    (6)     &       (7)      &        (8) 
         &      (9)     \\
\hline
                &     HE      &  $0.372 \pm 0.007$  & $0.106  \pm  0.021$ &  $7.9E-3 \pm  0.001$  &   3.486    &      688.1     &       562.5        &  
    6.61     \\
         1      &     ME      &  $0.368 \pm 0.008$  & $0.097  \pm  0.027$ &  $4.5E-2 \pm  0.009$  &   3.793    &      67.27     &       29.15        &     11.86    \\
	     
    &     LE      &  $0.356 \pm 0.010$  & $0.139  \pm  0.031$ &  $6.8E-2 \pm  0.013$  &   2.557    &      65.40     &       13.08        &     14.63    \\
\hline

                &     HE      &  $0.389 \pm 0.011$  & $0.118  \pm  0.036$ &  $9.7E-3 \pm  0.002$  &   3.274    &      639.4     &       547.5        &   
  7.89     \\
         2      &     ME      &  $0.424 \pm 0.021$  & $0.137  \pm  0.088$ &  $5.3E-2 \pm  0.018$  &   3.085    &      68.48     &       25.90        &     14.74    \\
      
           &     LE      &  $0.392 \pm 0.024$  & $0.117  \pm  0.018$ &  $8.5E-2 \pm  0.013$  &   3.344    &      73.54     &       13.00        &     14.72    \\
 \hline
 \end{tabular}
	\vspace{0.5cm}
\end{table*}

\begin{table*}[!h]
	\vspace{0.5cm}
\scriptsize
 \addtolength{\tabcolsep}{-1.0pt}
 \centering
\caption{QPO Properties from energy-dependent PDS using Insight-HXMT light curve in HE band on May 
2021 05 (Observation ID: P030401400).} 
 \label{tab:table4}
 \begin{tabular}{cccccccccc}
 \hline
 Exposure ID  &  Energy Range   &     $\nu_{qpo}$     &        FWHM         &       Norm           & $Q$-value  & $\sigma$ & Source Count  &   Background Count   &    RMS (\%)   \\  
    (1)       
 &       (2)       &          (3)        &         (4)         &       (5)            &    (6)     &    (7)   &      (8)      &   
       (9)          &      (10)     \\
\hline
              &     27-35       &  $0.388 \pm  0.009$ & $5.8E-2 \pm  0.027$ & $2.3E-2 \pm  0.006$  &   6.689    &   3.83   &     148.1     &       
  97.36         &      7.58     \\
              &     35-48       &  $0.379 \pm  0.009$ & $6.7E-2 \pm  0.026$ & $4.0E-2 \pm  0.010$  &   5.656    &   4.00   &     103.7     &        63.09     
    &      10.43    \\
      1       &     48-67       &  $0.373 \pm  0.010$ & $4.9E-2 \pm  0.021$ & $1.5E-2 \pm  0.005$  &   7.612    &   3.00   &     133.8     &        102.4         &    
   5.99     \\
              &     67-100      &  $0.373 \pm  0.010$ & $4.9E-2 \pm  0.042$ & $1.6E-2 \pm  0.008$  &   7.612    &   3.00   &     100.6     &        92.25         &      6.727    \\
 
              &     100-150     &          -          &         -           &         -            &     -      &    - 
     &     69.52     &        63.93         &        -      \\
              &     150-200     &          -          &         -  
          &         -            &     -      &    -     &     92.96     &        93.86         &        -      \\
\hline

              &     27-35       &  $0.405 \pm  0.008$ & $4.9E-2 \pm  0.028$ & $2.9E-2 \pm  
0.014$  &   8.265    &   2.07   &     145.4     &        104.5         &      8.12     \\
              &     35-48       &  $0.406 \pm  0.008$ & $5.1E-2 \pm  0.029$ & $4.5E-2 \pm  0.019$  &   7.960 
   &   2.37   &     100.3     &        60.59         &      9.63     \\
      2       &     48-67       &  $0.405 \pm  0.004$ & $4.6E-2 \pm  0.009$ & $3.9E-2 \pm  0.015$  &   8.804    &   2.60 
  &     128.7     &        87.59         &      8.92     \\
              &     67-100      &        0.415        &        4.6E-2       &    
   1.2E-2         &   9.020    &    -     &     104.6     &        98.45         &      5.71     \\
              &     100-150     &          
-          &         -           &         -            &     -      &    -     &     70.15     &        65.85         &  
       -      \\
              &     150-200     &          -          &         -           &         -            &   
  -      &    -     &     88.97     &        82.75         &        -      \\

 \hline
 \end{tabular}
	\vspace{0.5cm}
\end{table*}

Using the HE band data, we created $0.01$~s time-binned light curve into six different energy bands (27-35, 35-48, 48-67, 67-100, 100-150, \& 150-200 keV) to search for QPO energy dependence.
We produced PDS using those light curves, which are shown in Figure 8. Around 4 Hz, there was the presence of the fundamental QPO.
However, this nature was not present for all the six energy
bands.
We find that QPO nature was seen up to 100 keV, above which there was no QPO.
In the 67-100~keV energy band, the QPO was more prominent in epoch 1 than epoch 2. The data in epoch 2 
was more noisy and thus, we could not estimate the error from model fitting.
For both the epochs, the QPO nature was strongest in the 35-48 keV energy band, above which the nature started to 
weaken.
Also, in the 27-35 keV, the QPO nature was weaker than that in 35-48 keV.
This energy dependence gives a clear idea about the origin of QPO, which we will discuss later in the discussion
section.
The QPO properties from energy-dependent light curves are listed in Table 4. Along with $Q$-value and RMS, we also estimated the significance ($\sigma$) of the QPOs in these energy 
bands.
This is estimated as $\sigma = \frac{Norm}{err_{Norm}}$ (Sreehari et al. 2019), where $err_{Norm}$ is the `-'ve error of the Lorentzian model normalization ($LN$).
We also estimated 
the RMS from these properties, which is shown in the column 10. From the significance ($\sigma$) and RMS, we find that QPO was strongest in the 35-48 keV energy band for both the epochs, as 
just mentioned above.
We also searched for QPO using NuSTAR and NICER light curves on 2021 June 17 (MJD 59382).
However, no presence of QPO was found on that day.
\vskip 0.5cm
\subsection{Spectral Properties}

\begin{figure}[!h]
\centering
Unfolded Model Fitted Spectra\par\medskip
\vbox{
\includegraphics[width=5.5truecm,angle=270]{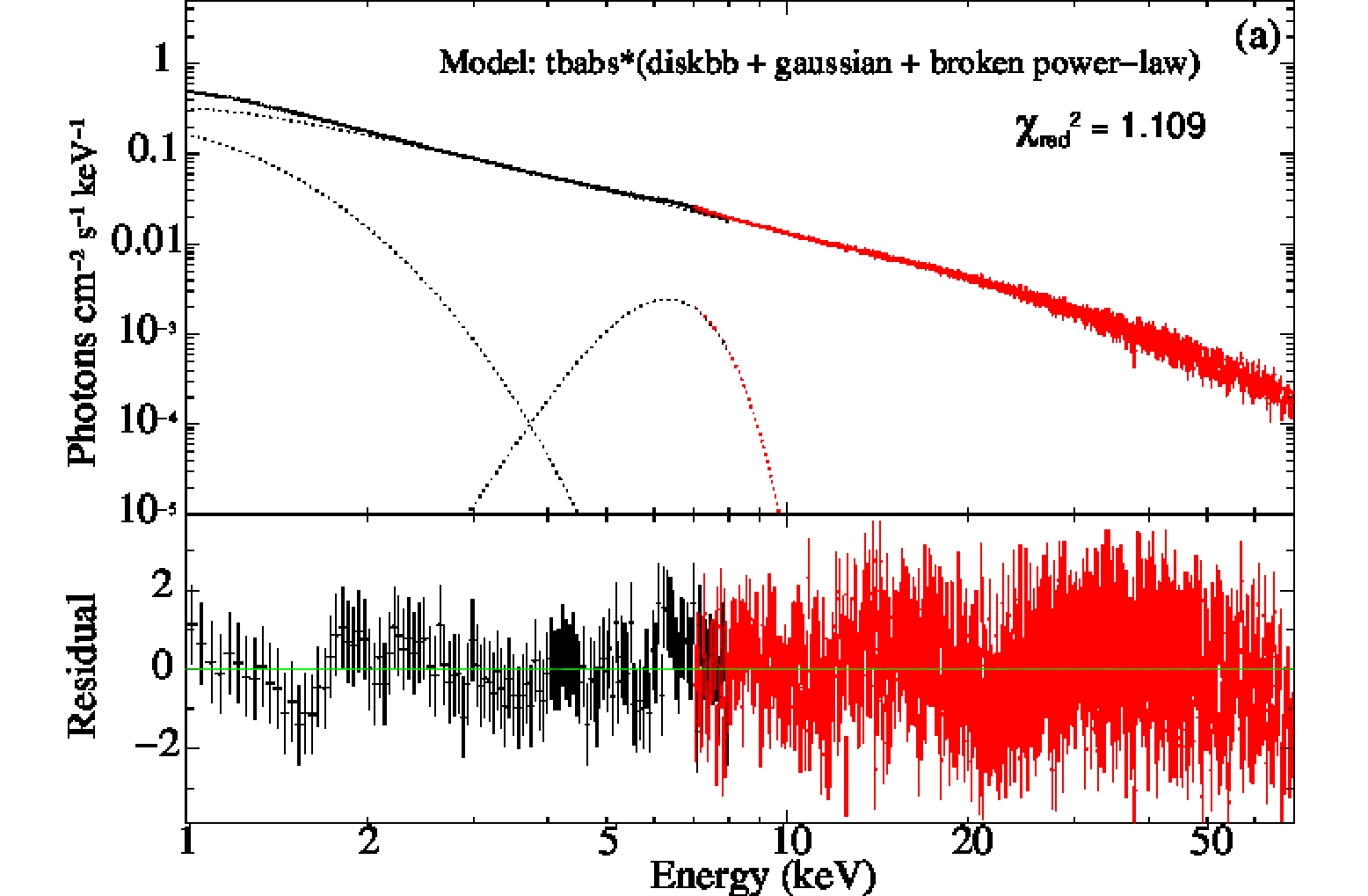}
\includegraphics[width=5.5truecm,angle=270]{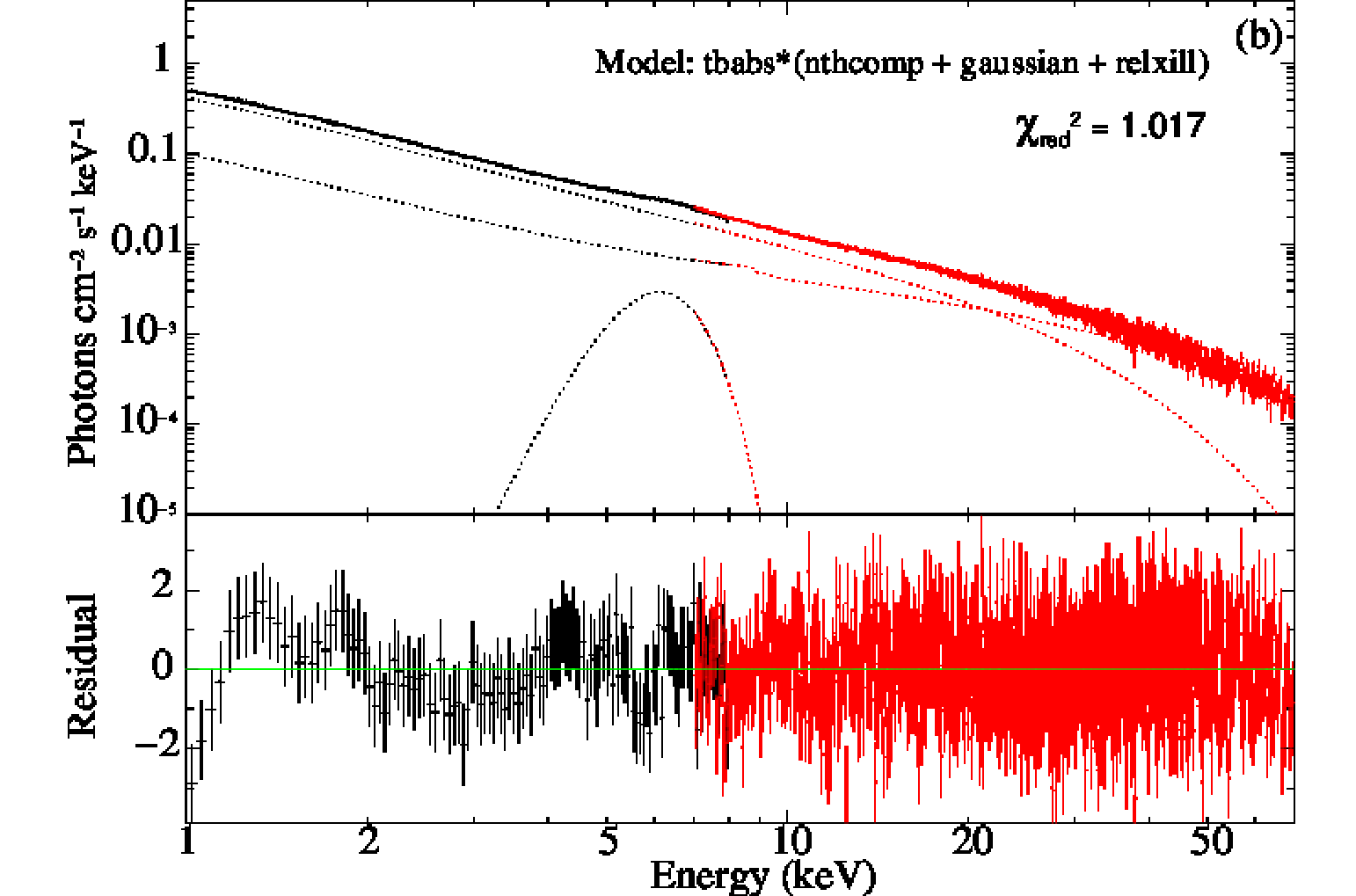}
\includegraphics[width=5.5truecm,angle=270]{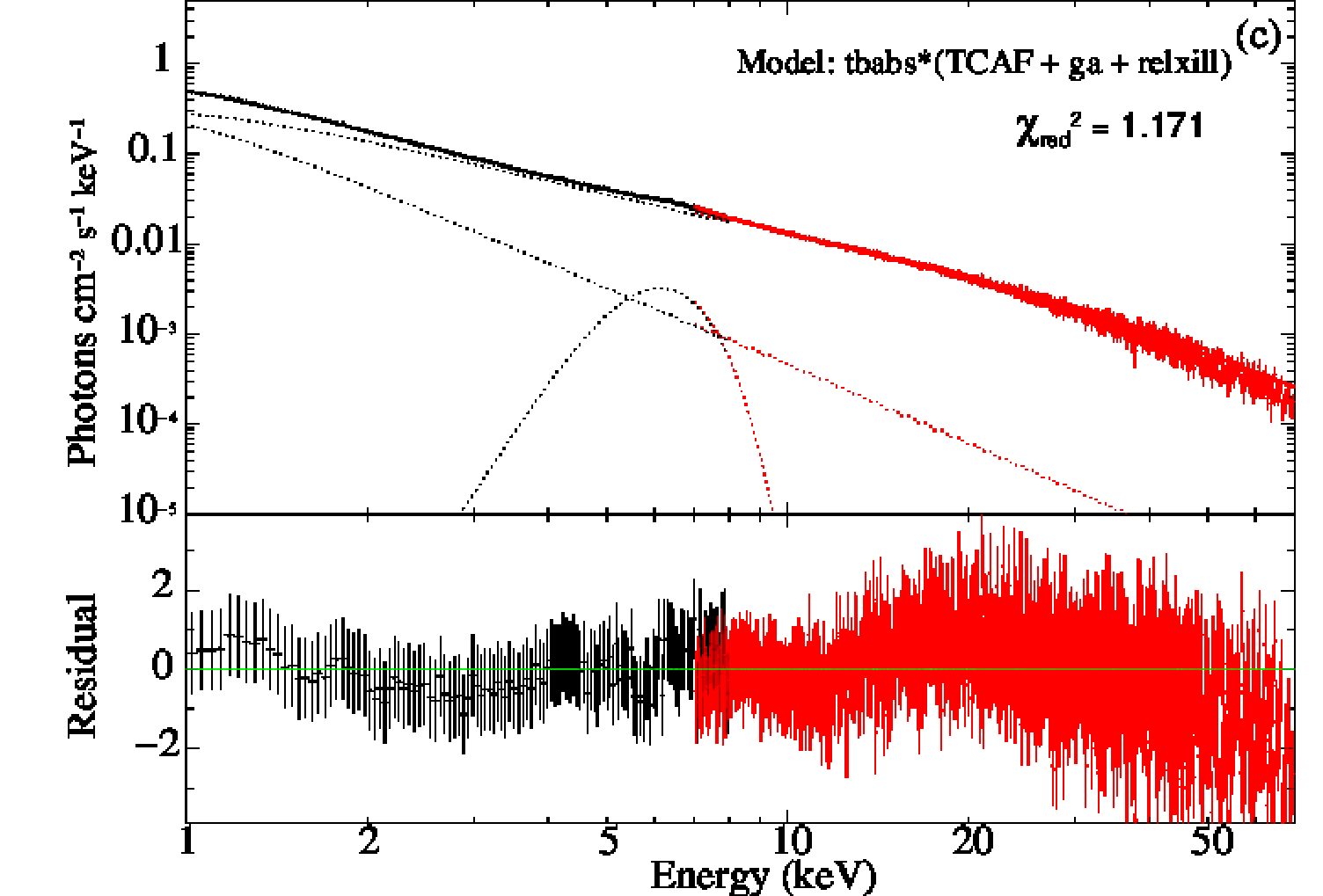}}
\caption{Model fitted unfolded broadband spectra in $1-70 ~keV$ energy band using \textit{NICER+NuSTAR} data.
(a), (b), and (c) show different sets of model combinations that fit the data 
         best.
This is for the obs. ID. 90702316002.}  
\end{figure}

A clear understanding of the spectral nature and evolution during an outburst can be obtained by examining the broadband spectra.
As described in Section 2.3, we spectrally 
fitted the simultaneous NICER and NuSTAR data in the broad range of 1–70 keV for both observation epochs (MJD 59339 and 59382) using various phenomenological and physical models.
In 
Figure 9, we show the best model-fitted unfolded spectra in the broadband 1–70 keV range using the three primary broadband model combinations.
All of our spectrally fitted parameters 
from this analysis are given in Table 5. We discuss the physical implications of these results below.

\vskip 0.2cm
\noindent{\bf Epoch-1 (2021 May 05 ; MJD 59339):}
Although the $n_H$ was consistent for the 1st and 4th model combination, it was deviated when using the 2nd model.
However, this is still in good agreement when it comes to hydrogen 
column density variation of stellar mass black holes.
The innerdisk temperature ($T_{in}$) was $\sim 0.31$~keV with $\Gamma1 \sim 1.62$ and $\Gamma2 \sim 2.43$.
The $\Gamma2$ takes 
care of the change of slope in higher energies for the phenomenological \textit{broken power-law} model, where in the physical model approach, the \textit{relxill} model takes care 
of that.
This is mainly thought to be present due to reflection.
In the 2nd model, $\Gamma$ was very close to the $\Gamma1$ of the first model.
This suggests that the source was in 
harder (HS or HIMS) state during this time.
We found the $Fe-K_\alpha$ line has a peak $\sim$ 6.14-6.32~keV from all the models implemented for Epoch-1.
The spin and inclination of 
the source was consistent from the 2nd and 4th models in the Table 5. We find that the spin and inclination were 0.993-0.998 and (70-76)$^\circ$ from both these models.
From the TCAF 
model parameters, we notice that the disk rate ($\dot{m}_d$) and halo rate ($\dot{m}_h$) were $\sim$ 0.056 and 0.032 respectively, whereas the shock location ($X_s$) was $\sim 130~r_s$ 
with compression ratio ($R$) $\sim 2.38$.
This also supports that the source was in harder state during this time, more suitably in hard intermediate state (HIMS).
From this model, 
we also estimated the mass of the source as $\sim 12 M_\odot$.
It is important to note that the reliability of this mass estimate depends on the assumptions built into the combined spectral model.
In our TCAF + relxill fit, several 
reflection parameters were frozen (e.g., $Index1=3$, $Index2=3$, $R_{in}=1~R_{ISCO}$, and $E_{cut}=300$ keV).
The high-energy cutoff could not be constrained as it lies far beyond 
our NuSTAR upper energy limit of 70 keV, and the emissivity indices were frozen to standard disk values to prevent model degeneracies.
While freezing these parameters is 
necessary to achieve a stable fit, it restricts the shape of the reflection continuum.
Consequently, any unmodeled complexities in the reflection geometry could force the TCAF 
continuum to shift to compensate, slightly altering the inferred flow parameters ($\dot{m}_d$, $\dot{m}_h$) and the model normalization.
Therefore, the derived black hole mass 
of $\sim 12 M_{\odot}$ should be viewed with the understanding that it carries an inherent systematic uncertainty strictly tied to these reflection geometry assumptions.

\vskip 0.2cm
\noindent{\bf Epoch-2 (2021 June 17 ; MJD 59382):}
In this epoch, the hydrogen column density ($n_H$) varied between (0.32-0.34) $\times 10^{22}$ cm$^{-2}$.
The innerdisk temperature got higher with a value of $\sim 0.84$~keV.
Spectral 
slopes also changed as we notice $\Gamma1$ and $\Gamma2$ are $\sim 3$ and $1.1$ respectively from first model in the table.
From the 3rd model in the table, we find that $\Gamma$ from 
\textit{relxill} model was $\sim 2.9$.
This means that the source made a transition towards softer (SIMS or SS) state.
For this epoch, we also find (from 3rd model) that the spin and
inclination was consistent with the epoch before with values of $a \sim 0.993$ and $i \sim 77^\circ$, respectively.
\begin{table}
\scriptsize
 \addtolength{\tabcolsep}{-0.5pt}
 \centering
\caption{Parameters from Broadband Spectral Fitting using NICER (1-8 keV) + NuSTAR (7-70 keV).
In column 1, we list the models that we used for our spectral analysis.
In column 2, we list 
	 the parameters of the models in column 1. Column 3, \& 4 represent the values of those respective parameters from our spectral analysis in Epoch-1 and Epoch-2 respectively.
The 
	 value of the parameter $KT_{bb}$ of the \textit{nthcomp} model was set to the $T_{in}$ of the \textit{diskbb} model for the first epoch.
For \textit{relxill} model we assumed some 
	 values and freeze them while doinf the spectral fitting.
These are: $Index1=3.0$, $Index2=3.0$, $Rbr=15$, $R_{in}=1.0$, $R_{out}=1000$, $z=0$, $E_{cut}=300$. In case of the 2nd 
	 model, the $\Gamma$ of the \textit{relxill} model was fixed to the value of $\Gamma$ of the \textit{nthcomp} model.
$n_H$ is in the units of 10$^{22}$~cm$^{-2}$.} 
 \label{tab:table5}
 \begin{tabular}{ccccc}
 \hline
 Models  &   Parameters    &          Epoch-1           &          Epoch-2           \\
  (1)    &      (2)        &            (3)      
        &            (4)             \\
\hline
         &     $n_H$       &   $  0.300 \pm 7.1E-2   $  &   $  0.3409 \pm 1.9E-2  $  \\
         &    $T_{in}$     &   $ 0.3089 \pm 
1.2E-2   $  &   $  0.8421 \pm 1.7E-3  $  \\
         &     Norm        &   $ 4971.8 \pm 1603.4   $  &   $   739.7 \pm 7.87    $  \\
  DBB    &    $\Gamma_1$   &   $ 1.6209 \pm 4.8E-3   $  &   $   3.017 \pm 0.136   $  
\\
  $+$    &   $E_{break}$   &   $ 22.236 \pm 0.1772   $  &   $  18.994 \pm 0.949   $  \\
  BKNPL  &    $\Gamma_2$   &   $ 2.4271 \pm 1.4E-2   $  &   $  1.0626 \pm 0.1462  $  \\
  $+$    &     Norm        &   $ 0.5394 \pm 5.2E-3   $  &  
  $  0.1802 \pm 6.4E-2  $  \\
  Ga	 &    $E_{peak}$   &   $ 6.3158 \pm 9.2-2    $  &   $         -           $  \\
	 &    $\sigma$     &   $ 1.0222 \pm 6.4E-2   $  &   $         -           
$  \\
	 &      Norm       &   $ 6.3E-3 \pm 5.6E-4   $  &   $         -           $  \\
         & $\chi^2/DOF$    &   $ 1359.5/1225         $  &   $      260.3/265      $  \\
\hline  
                                                                            
         &     $n_H$       &   $ 
5.7E-2 \pm 1.2E-2   $  &   $         -           $  \\
         &   $\Gamma$      &   $ 1.6525 \pm 1.7E-2   $  &   $         -           $  \\
         &    $KT_e$ 
       &   $   8.58 \pm 0.34     $  &   $         -           $  \\
         &    $Norm$       &   $  0.298 \pm 2.9E-2   $  &   $         -          
  $  \\
 nthcomp &      $a$        &   $  0.993 \pm 0.128    $  &   $         -           $  \\
 $+$     &    $\theta$     &   $  76.05 \pm 1.601    $  &   $         -   
         $  \\
 relxill &    $log{xi}$    &   $  3.514 \pm 0.275    $  &   $         -           $  \\
 $+$     &      $Afe$      &   $  0.523 \pm 5.6E-2   $  &   $         
-           $  \\
 Ga	 &  $refl_{frac}$  &   $  10.23 \pm 0.164    $  &   $         -           $  \\
	 &    $Norm$       &   $ 1.6E-3 \pm 6.6E-4   $  &   $         -        
   $  \\
         &   $E_{peak}$    &   $ 6.1362 \pm 0.116    $  &   $         -           $  \\
         &   $\sigma$      &   $ 0.9176 \pm 8.2E-2   $  &   $         - 
          $  \\
         &     Norm        &   $ 6.5E-3 \pm 7.2E-4   $  &   $         -           $  \\
         &  $\chi^2/DOF$   &   $   1370.3/1222       $  & 
  $         -           $  \\
\hline                                                                         
   
      &     $n_H$       &   $       -             $  &   $  0.3254 \pm 3.4E-2  $  \\
         &    $T_{in}$     &   $       -             $  &   
$  0.8436 \pm 3.7E-3  $  \\
         &     Norm        &   $       -             $  &   $   735.3 \pm 16.8    $  \\
         &      $a$        &   $     
  -             $  &   $   0.993 \pm 6.9E-2  $  \\
  DBB    &    $\theta$     &   $       -             $  &   $   77.16 \pm 5.86    $  \\
  $+$    &    $\Gamma$     &  
 $       -             $  &   $   2.905 \pm 0.286   $  \\
 relxill &    $log{xi}$    &   $       -             $  &   $   1.712 \pm 0.094   $  \\
         &      $Afe$ 
      &   $       -             $  &   $   5.938 \pm 0.312   $  \\
         &  $refl_{frac}$  &   $       -             $  &   $  4.1356 \pm 0.46    $  \\
      
   &    $Norm$       &   $       -             $  &   $  2.2E-3 \pm 6.3E-5  $  \\
         &  $\chi^2/DOF$   &   $       -             $  &   $     277.11/263   
    $  \\
 \hline                                                                            
         &     $n_H$    
   &   $ 0.266  \pm 7.8E-3   $  &   $          -          $  \\
         & ${\dot m}_d$    &   $ 5.6E-2 \pm 5.0E-3   $  &   $          -          $  \\
         
 & ${\dot m}_h$    &   $ 3.2E-2 \pm 3.0E-3   $  &   $          -          $  \\
         &    $X_s$        &   $  129.5 \pm 7.7      $  &   $          -         
  $  \\
         &     $R$         &   $   2.38 \pm 0.23     $  &   $          -          $  \\
	 &  $M_{BH}$       &   $  11.95 \pm 0.26     $  &   $       
    -          $  \\
 TCAF    &   $Norm$        &   $  156.3 \pm 2.5      $  &   $          -          $  \\
 $+$     &  $E_{peak}$     &   $  6.15  \pm 8.8E-2   $  &   
$          -          $  \\
 Ga      &  $\sigma$       &   $  0.97  \pm 7.6E-2   $  &   $          -          $  \\
 $+$     &    Norm         &   $ 7.9E-3 
\pm 6.1E-4   $  &   $          -          $  \\
 relxill &      $a$        &   $ 0.998  \pm 1.5E-2   $  &   $          -          $  \\
         &    $\theta$   
  &   $ 70.0   \pm 0.31     $  &   $          -          $  \\
         &    $\Gamma$     &   $ 1.61   \pm 1.7E-3   $  &   $          -          $  \\
  
       &    $log{xi}$    &   $   4.7  \pm 5.9E-2   $  &   $          -          $  \\
         &      $Afe$      &   $   0.5  \pm 0.05     $  &   $      
     -          $  \\
         &  $refl_{frac}$  &   $ 10.11  \pm 4.4E-2   $  &   $          -          $  \\
         &    $Norm$       &   $ 1.8E-3 \pm 1.23e-5  $  &   $ 
         -          $  \\
         &  $\chi^2/DOF$   &   $    1426.37/1218     $  &   $                     $  \\
 \hline
 \end{tabular}
 \vspace{0.5cm}
\end{table}

\section{Discussions} 

In this work, we revisited the 2021 outburst of MAXI J1803--298 to investigate the physical properties of the inner accretion flow 
using strictly simultaneous multi-instrument 
data. Rather than solely re-classifying the well-known spectral states of this outburst, our analysis focused on extracting novel physical constraints.
Specifically, we utilized the 
TCAF model to independently estimate the black hole mass, extended the detection of the LFQPO into the hard X-ray regime ($\sim 100$ keV) using \textit{Insight-HXMT}, and investigated 
the rapid variations of the LFQPO frequency in \textit{NuSTAR} observation with dynamic cospectral analysis.
We discuss the physical implications of these specific findings below.

\subsection{LFQPO characteristics and physical origin}

The 2021 outburst of the black hole candidate MAXI~J1803--298 provides an excellent opportunity to investigate the origin of low-frequency quasi-periodic oscillations (LFQPOs) using
simultaneous broadband spectral and timing observations.
During the early epoch (2021 May 05), a clear LFQPO is detected in the \textit{NuSTAR} observation.
Fast timing studies 
of bright Galactic X-ray binaries with \textit{NuSTAR} are complicated by the instrument's long ($\sim 2.5$ ms) and variable dead time.
This dead time alters the recorded photon count 
rate and introduces spurious correlations between originally independent events, resulting in a distorted power density spectrum (PDS) where the white noise level does not remain constant.
As seen in Figure 3, the Leahy-normalized PDS from both FPMA and FPMB modules dip below the expected Poisson noise level of 2 above $\sim 1$ Hz and exhibit broad wave-like features.
To mitigate these systematic effects, we employed the analysis technique proposed by Bachetti et al.
(2015) which takes advantage of the two independent detector units of \textit{NuSTAR}.
By calculating the real part of the complex cross-spectrum (the \textit{cospectrum}) between the independent FPMA and FPMB light curves, effects of non-correlated variability in the 
lightcurves, such as distortions in the white noise level due to dead-time, is effectively suppressed.
This approach provides a reliable estimate of the intrinsic source variability 
and eliminates the need for complex modeling of the variable dead time.
While the cospectrum has been widely utilized as a proxy for the white noise subtracted PDS in various timing 
studies (Ingram et al. 2016, Nathan et al. 2022), we have also generated the dynamic cospectrum (see Figure 6).
From Figure 6, we can see the centroid frequency of the detected LFQPO 
has increased rapidly from $\sim$0.35~Hz to $\sim$0.5~Hz within the single \textit{NuSTAR} observation (also see Figure 5).

The detection of such rapid frequency evolution on intra-observational timescales indicates that the inner accretion flow is dynamically evolving rather than maintaining a static 
configuration.
This rapid evolution in the physical size or structure of the Comptonizing region can naturally drive the observed changes in the QPO frequency.
This behavior is consistent 
with both intrinsic dynamical instabilities (such as oscillating shocks) and geometric models where the precession frequency depends on the evolving scale of the inner flow (e.g., 
Lense-Thirring precession; Ingram et al. 2009).

The LFQPO is detected over a broad energy range, extending up to $\sim$100~keV in the \textit{Insight-HXMT} data (see Fig. 8).
At these energies, the contribution from the optically 
thick accretion disk is negligible, implying that the modulation originates primarily within the Comptonized emission region.
This interpretation is further supported by the energy 
dependence of the fractional rms variability, which peaks in the 35--48~keV band and decreases toward both lower and higher energies.
The absence of a detectable QPO above 
$\sim$100~keV must be interpreted with caution.
As shown in Table 4, the source count rates in the 100--150~keV and 150--200~keV bands drop significantly, becoming comparable to or 
even lower than the background levels.
This results in a poor signal-to-noise ratio that makes the detection of any QPO signal extremely difficult, regardless of its intrinsic strength.
Consequently, we cannot rule out that the non-detection at the highest energies is primarily a result of limited photon statistics.
However, if the QPO amplitude is intrinsically 
suppressed at these energies, it could suggest that coherent modulation is no longer efficiently preserved, possibly due to increased multiple scattering or reduced coherence in the 
hottest, innermost regions of the flow.
Future high-sensitivity hard X-ray observations are required to disentangle these statistical and physical effects.

\subsection{Accretion geometry from spectral--timing analysis}

Broadband spectral modeling of the early epoch indicates a hard or hard-intermediate accretion state dominated by Comptonized emission, together with the presence of reflection features.
Within the framework of the two-component advective flow (TCAF) model, the inferred accretion geometry consists of a sub-Keplerian halo and a weaker Keplerian disk, with a shock located 
at $\sim$130 Schwarzschild radii.
The characteristic dynamical timescale associated with this shock location is consistent with the observed LFQPO frequency, suggesting that oscillations 
of the post-shock Comptonizing region may contribute to the observed variability.
Although the present analysis does not uniquely discriminate among different LFQPO models, the consistency 
between the timing properties and the inferred accretion geometry supports an origin linked to the dynamics of the inner hot flow rather than the optically thick disk.
In contrast, no LFQPO is detected during the later epoch (2021 June 17), when the source exhibits a softer spectrum characterized by a higher disk temperature and a steeper photon index.
This behavior is consistent with the canonical evolution of black hole X-ray binaries, in which LFQPOs are suppressed as the source transitions toward softer accretion states and the 
relative contribution of the Comptonizing region diminishes.
The comparison between the two epochs therefore reinforces the close connection between the presence of LFQPOs and the physical 
extent of the inner hot flow.
An additional outcome of the TCAF spectral modeling is the constraint obtained on the black hole mass.
The spectral fits yield a mass estimate consistent with a stellar-mass black hole, 
providing an independent confirmation of the nature of MAXI~J1803--298 as a Galactic black hole X-ray binary.
The inferred mass range is physically consistent with the observed LFQPO 
frequencies and the estimated shock location, since the characteristic dynamical timescale of the post-shock Comptonizing region depends explicitly on the black hole mass.
The consistency 
between the TCAF-derived mass and the observed timing properties therefore supports the interpretation that the detected LFQPO originates from oscillations or dynamical modulation of the 
inner hot accretion flow.
Although mass estimates obtained through spectral modeling may carry systematic uncertainties related to model assumptions and parameter degeneracies, the agreement 
between the spectral and timing constraints strengthens confidence in the inferred accretion geometry and the physical interpretation of the variability observed in MAXI~J1803--298.
\subsection{Inclination effects and timing methodology}

MAXI~J1803--298 is a high-inclination system and displays dipping behavior during the early phase of the outburst, suggesting that geometric effects such as variable absorption or partial 
obscuration may influence the observed emission.
However, the detection of coherent LFQPOs across multiple instruments and over a wide energy range confirms that the primary 
modulation mechanism is deeply rooted in the inner Comptonizing region, rather than being an artifact of outer-disk obscuration or localized line-of-sight absorption.
This broad-band 
nature is consistent with scenarios where the entire inner hot flow modulates the X-ray flux, whether through intrinsic dynamical variations (such as shock oscillations in the TCAF 
framework) or through the geometric Lense-Thirring precession of the hot flow itself.

From a methodological perspective, this work highlights the importance of employing appropriate timing techniques for \textit{NuSTAR} observations of bright Galactic X-ray binaries.
The 
long and variable instrumental dead time of \textit{NuSTAR} can significantly distort conventional power density spectra, particularly at high count rates.
By constructing the cospectrum 
from the independent FPMA and FPMB detectors, non-correlated noise components, including dead-time effects, are effectively suppressed, enabling robust detection of intrinsic LFQPO features.
This approach strengthens confidence in the timing results and demonstrates the utility of cross-detector techniques for broadband variability studies.
Overall, the combined spectral and timing results support a scenario in which the observed LFQPO originates from the Comptonizing inner accretion flow, whose physical properties evolve 
during the outburst.
Although limited to two epochs with strictly simultaneous coverage, the present study provides a coherent observational framework that motivates future investigations 
employing phase-resolved spectroscopy and time-lag analysis (check Ingram et al. 2015, 2016; Shui et al. 2023) to further constrain the geometry and dynamics of the innermost accretion 
regions in MAXI~J1803--298.
\section{Summary and Conclusions}

We present a broadband spectral and timing study of the black hole candidate MAXI~J1803--298 during its 2021 outburst using simultaneous observations from NICER, \textit{NuSTAR}, and 
\textit{Insight-HXMT}.
The multi-instrument coverage allows us to investigate the evolution of low-frequency quasi-periodic oscillations (LFQPOs) together with the spectral properties 
of the source over a wide energy range.
During the early observation epoch, the source exhibits a hard or hard-intermediate spectral state dominated by Comptonized emission along with reflection features.
Spectral modeling 
within the framework of the two-component advective flow (TCAF) model indicates the presence of a sub-Keplerian halo and a Keplerian disk with a shock located at $\sim$130 Schwarzschild 
radii.
The TCAF modeling also provides an independent estimate of the black hole mass of $\sim$ 12 $M_\odot$, in agreement with the observed LFQPO timescales.
A prominent LFQPO is detected in the \textit{NuSTAR} observation during this early epoch, whose centroid frequency is observed to be rapidly evolving from $\sim$0.35~Hz to $\sim$
0.5~Hz in the dynamic cospectra within the single observation.
The LFQPO is prominently detected upto an energy of $\sim$100~keV in the \textit{Insight-HXMT} observation. The energy 
dependence of the fractional rms variability indicates that the modulation originates primarily from the Comptonizing inner accretion flow.
In contrast, no LFQPO is detected during the 
later observation epoch, when the source exhibits a softer spectral state characterized by stronger disk emission and a steeper photon index, consistent with the suppression of rapid 
variability during state transitions.
Overall, the combined spectral and timing results support a scenario in which LFQPOs in MAXI~J1803--298 arise from the dynamically evolving Comptonizing inner accretion flow.
Future 
investigations involving phase-resolved spectroscopy, time-lag analysis, and long-term monitoring will be essential for further constraining the physical mechanisms responsible for QPO 
production and accretion flow evolution in this system.
\section{Data Availability}

This work has made use of public data from several satellite/instrument archives and has made use of software from the HEASARC, which is developed and monitored by the Astrophysics 
Science Division at NASA/GSFC and the High Energy Astrophysics Division of the Smithsonian Astrophysical Observatory.
NICER data is provided by NASA/GSFC. FPM data from the NuSTAR 
mission is led by Caltech, which is funded by NASA and managed by NASA/JPL.
The NuSTARDAS software package is jointly developed by ASDC, Italy and Caltech, USA.
The {\it Insight}-HXMT 
mission is funded by the China National Space Administration (CNSA) and the Chinese Academy of Sciences (CAS).
\section{Acknowledgements}

We thank the scientific editor for assessing the manuscript in a timely manner.
We acknowledge the anonymous referee(s) for his/her/their detailed comments and insightful suggestions 
that have improved the quality of the manuscript.
KC acknowledges support from the CaiYun-2024 Postdoctoral Fund for Innovation Project of Yunnan Province (C615300504128).

\bibliographystyle{elsarticle-harv}

\end{document}